\documentclass[preprint,12pt]{elsarticle}

\usepackage{amssymb}
\usepackage{bm}
\usepackage{graphicx}
\usepackage{miller}
\usepackage{subcaption}
\usepackage{mathrsfs}
\usepackage{dcolumn}
\usepackage{tabularx}
\usepackage[version=4]{mhchem}
\def\etal{\emph{et al}.}
\usepackage{color}

\newcommand*{\rtten}[1]{\bm{#1}}
\newcommand*{\rtvec}[1]{\mathbf{#1}}

\newcommand{\vv}[1]{{\textbf{\textcolor{red}{Venkat: #1}}}}

\graphicspath{{Figures/}}

\begin{document}

\begin{frontmatter}



\title{Chemomechanics: friend or foe of the ``AND problem'' of solid-state batteries?}


\author[inst1,inst2]{Zeeshan Ahmad}

\affiliation[inst1]{organization={Department of Mechanical Engineering, Carnegie Mellon University},
            city={Pittsburgh},
            postcode={15213}, 
            state={PA},
            country={United States}}

\author[inst1]{Victor Venturi}
\author[inst1]{Shashank Sripad}
\author[inst1]{Venkatasubramanian Viswanathan}
\ead{venkvis@cmu.edu}

\affiliation[inst2]{organization={Pritzker School of Molecular Engineering, University of Chicago},
            city={Chicago},
            postcode={60637}, 
            state={IL},
            country={United States}}

\begin{abstract}
Solid electrolytes are widely considered as the enabler of lithium metal anodes for safe, durable, and high energy density rechargeable lithium-ion batteries. Despite the promise, failure mechanisms associated with solid-state batteries are not well-established, largely due to limited understanding of the chemomechanical factors governing them. We focus on the recent developments in understanding solid-state aspects including the effects of mechanical stresses, constitutive relations, fracture, and void formation, and outline the gaps in the literature. We also provide an overview of the manufacturing and processing of solid-state batteries in relation to chemomechanics. The gaps identified provide concrete directions towards the rational design and development of  failure-resistant solid-state batteries.
\end{abstract}




\end{frontmatter}


\section{Introduction}
\label{sec:intro}
Solid-state batteries (SSBs) based on lithium (Li) metal anodes present a promising opportunity to enhance the energy density, stability, longevity, and reliability of current Li-ion batteries~\cite{manthiram2017solid,Robinson2014,janekSolidFutureBattery2016a}. These batteries could provide the energy and power densities required for long-range and fast charge and discharge applications, for example, in electric vehicles and electric aircraft~\cite{fredericksPerformanceMetricsRequired2018,moore2014aviation,billsPerformanceMetricsRequired2020}.  Solid electrolytes  must satisfy many constraints simultaneously for use in Li metal batteries, popularly termed as the ``AND problem'', i.e. they must possess sufficient ionic conductivity~\cite{linDesignCationTransport2020a}, be stable to the lithium metal anode and the cathode~\cite{Xiao2019understanding,Wenzel2016direct,Banerjee2020interfaces}, handle large mechanical deformations at the interface~\cite{Zhang2017electro}, maintain interfacial contact and low resistance through cycling~\cite{Yu2017accessing}, be compatible with a scalable manufacturing approach to produce high-quality low defect material~\cite{schnell2018all}, among other requirements. While there are solid electrolytes that can satisfy one or two criteria, the challenge is trying to satisfy all of them simultaneously with a single material.

The ionic conductivity limitations, which were initially a major bottleneck for the adoption of solid electrolytes in batteries, have largely been resolved. Numerous classes of fast solid Li-ion conductors like thiophosphates~\cite{kamayaLithiumSuperionicConductor2011,Seino2014,Deiseroth2008}, antiperovskites~\cite{Zhao2012superionic,Zhang2013abinitio} and garnets~\cite{Thangadurai2003,Murugan2007} have been discovered with Li-ion conductivities even exceeding that of liquid electrolytes ($\sim$ 1 mS/cm) in some cases. This, coupled with low interfacial resistance, enables fast charging/discharging or high power applications with solid electrolytes~\cite{Kato2016-ssb}. Beyond this requirement, the solid electrolyte needs to have a stable interface with the Li metal anode and the cathode and maintain it during operation to ensure high Coulombic efficiency. This implies that a good solid electrolyte should be electrochemically and chemically stable against both cathode and anode or it should form stable decomposition products~\cite{Richards2015interface,Chen2019approaching}. Additionally, during charging and discharging, the solid electrolyte needs to suppress instabilities associated with plating and stripping Li metal, like dendrite growth, cracking, and pitting~\cite{Ren2015direct,porzMechanismLithiumMetal2017,kasemchainanCriticalStrippingCurrent2019}. Cost considerations further dictate that solid electrolytes that can be easily integrated and manufactured at scale are most likely to be commercialized~\cite{huang2021manufacturing}. Satisfying these criteria simultaneously requires a fundamental understanding of chemomechanical processes involved with solid electrolytes and the associated interfaces~\cite{Lewis2019chemo}.


Here, we begin by summarizing the key properties of interest related to SSBs and identify their links to various failure mechanisms. We first highlight the advances in our understanding of solid-state electrochemomechanics involving the mechanical properties, constitutive laws, external device parameters, such as stack pressure, and their effects on the thermodynamics and kinetics of electrodeposition and stripping. We then review some of the major failure mechanisms and their fundamentals from a chemomechanical point of view. For each failure mechanism, we provide areas where there are gaps in the understanding and how research in those areas can help mitigate the corresponding problem. We conclude the review with a summary of the failure mechanisms discussed and an outlook for the research that can shape our solutions to those challenges.


\section{Key Features}
\begin{figure}[htbp]
    \centering
    \includegraphics[width=1\textwidth]{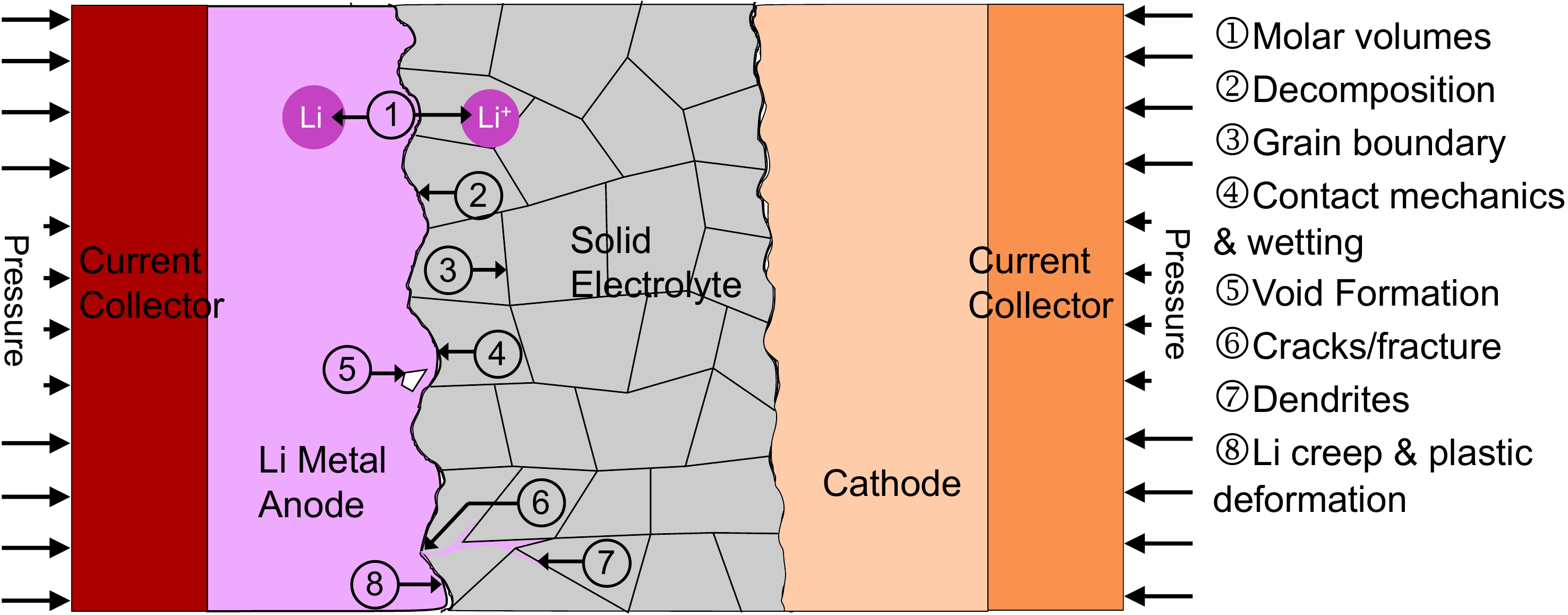}
    \caption{Schematic of a Li metal SSB along with the key chemomechanical phenomena and challenges at the Li metal-solid electrolyte interfaces. The interface is characterized by several new phenomena like contact loss, wetting, void formation, the molar volume change of Li, surface defects, and electrolyte decomposition. The solid electrolyte may have cracks at the surface and grain boundaries within. Key failure mechanisms are Li penetration or dendrite growth, cracking at the surfaces, and fracture and pitting.}
    \label{fig:schematic}
\end{figure}

Fig.~\ref{fig:schematic} shows a schematic of a Li metal anode-based SSB along with the chemomechanical features reviewed in this manuscript.

The hallmark property of solids is the development of stresses in response to deformations or strains. Mechanics plays a role in batteries with liquid electrolytes as well through the development of stresses at the solid electrodes and separators. However, due to the mechanical compliance of liquid electrolytes unlike solid electrolytes, other effects such as chemical, thermal, and electrical dominate the mechanical effects~\cite{zhaoReviewModelingElectrochemomechanics2019a}.
In this review, we will focus mainly on the novel chemomechanical phenomena encountered due to the presence of a solid electrolyte at the anode-electrolyte interface. We now highlight the major advances in our understanding of chemomechanical effects in SSBs.

\subsection{Thermodynamics \& the electrochemomechanical free energy}
The central thermodynamic quantities involved in the study of electrochemomechanical aspects of SSBs are the response of solids to applied stress or strain and the response of the free energy of the species involved to the stress. The response to applied stress or strain is defined by the constitutive relations generally written in terms of the Cauchy stress tensor $\rtten{\sigma}$ (MPa) and strain tensor $\rtten{\epsilon}$. The change in the free energy $G$ (J/mol) of a species due to a general stress tensor $\rtten{\sigma}$ is given by
\begin{equation}
    dG = \sum_{i,j}\frac{\partial G}{\partial \rtten{\sigma}_{ij}}\Bigr|_{T,c}  d\rtten{\sigma}_{ij}
\end{equation}
The stress tensor can be decomposed into hydrostatic and deviatoric components $\rtten{\sigma} = -p\rtten{I}+\rtten{\tau}$. The quantity $\partial G/ \partial \rtten{\sigma}_{ij}$ has units of m$^3$/mol and hence, it can be interpreted as a molar volume tensor which generalizes the concept of the molar volume $\Omega = \partial G/\partial p$ under hydrostatic pressure~\cite{goyalNewFoundationsNewman2017}. Various simplified forms of the free energy due to mechanical stresses can be derived varying from $\Delta G = \Omega p$ to more sophisticated forms $\Delta G = -\Omega \rtvec{n}^T \rtten{\sigma} \rtvec{n}$ involving the deviatoric components of the stress tensor and the normal vector $\rtvec{n}$ to the surface~\cite{Monroe2004Effect,ganserExtendedFormulationButlerVolmer2019,Tikekar2016}.

The mechanical contribution to the free energy or the electrochemomechanical potential can also be derived by writing down the second law of thermodynamics for the system and applying the elegant Coleman-Noll procedure~\cite{Coleman1963}. This form has been derived under certain assumptions, for example, by Bucci \etal~\cite{bucciFormulationCoupledElectrochemical2016} and Ganser \etal~\cite{ganserFiniteStrainElectrochemomechanical2019}.

\subsection{Kinetics}
The chemical reaction at the metal electrode surface is given by:
\begin{equation}
    \mathrm{M \rightleftharpoons M^{+} + e^{-}}
\end{equation}
where M is used to denote the metal. At thermodynamic equilibrium, the free energies of the individual species are related by: $G_M = G_{M^{+}}+G_{e^-}$. The connection between kinetics and thermodynamics comes from the modification of the energy barrier due to the change of reactant and product free energies via the Bronsted-Evans-Polanyi principle~\cite{ganserExtendedFormulationButlerVolmer2019}. In general, the effect of mechanical stresses on the transition state free energy $G_{TS}$ can be written as:
\begin{equation}
   G_{TS} = G_{TS}^{ref} + \delta^{ref}_M G_{M}^{ref} + \delta^{mech}_{M} G_M^{mech} + \delta^{ref}_{M^+} G_{M^+}^{ref} + \delta^{mech}_{M^+} G_{M^+}^{mech}
\end{equation}
where the scaling factors $\delta$ determine the scaling of the transition state energy with the energies of the reactants and products. For example, $\delta_M^{ref}$ represents the scaling of the energy of the transition state with the reference energy of the metal M and $\delta^{mech}_{M}$ represents the scaling of the transition state energy with the mechanical part of the energy of the metal M. Note that the barrier is assumed to be independent of the free energy contribution from the configurational entropy. Various values of $\delta^{mech}$ can be found in the literature which corresponds to different ways the mechanical part of free energy can affect the barrier. A common approach in the literature is to assume $\delta^{mech}_M=0$ and $\delta^{mech}_{M^+}=1$~\cite{Monroe2005Impact,ahmad2017stability,baraiLithiumDendriteGrowth2017,zhangPressureDrivenInterfaceEvolution2020,Ahmad2020design}, which implies that the barrier is only affected by the mechanical part of the free energy of the electrolyte. Setting both $\delta^{mech}_M$ and $\delta_{M^+}^{mech}$ to zero implies the barrier is unaffected by mechanics. Mechanics still modifies the overpotential through the change in the equilibrium potential difference between the electrode and electrolyte and hence, the reaction kinetics.

The current density $i$ (mA/cm$^2$) at the electrode-electrolyte interface is governed by the Butler-Volmer equation written in terms of the overpotential $\eta$ (mV) as:
 \begin{equation}\label{eq:bv}
    i=i_0\left[  \exp\left\{ (1-\alpha)\frac{ F\eta}{RT} \right\}
-  \exp \left\{ -\alpha\frac{ F\eta}{RT} \right\} \right]
    \end{equation}
where $T$ is the temperature in K, $F$ is the Faraday's constant equal to 96,485.33 C/mol, $\alpha$ is the symmetry factor, $R$ is the gas constant equal to 8.3145 J/(mol$\cdot$ K) and $i_0$ is the exchange current density given by
\begin{equation}
        i_0 = i_0^{ref} \exp \left[ \frac{(\alpha - \delta^{mech}_M)\Delta G^{mech}_M + (1-\alpha - \delta_{M^{+}}^{mech})\Delta G^{mech}_{M^+} }{RT} \right]
    \end{equation}
where $\Delta G^{mech}$ is the contribution of the mechanics to the free energies~\cite{ganserExtendedFormulationButlerVolmer2019,klinsmannDendriticCrackingSolid2019} and $i_0^{ref}$ is the exchange current density without mechanical effects. Setting $\delta^{mech}_M=0$ and $\delta^{mech}_{M^+}=1$ and using $\Delta G_M^{mech}=\Delta G_{M^+}^{mech}  + \Delta G^{mech,eqbm}_{e^-}$ in this equation gives the commonly used relation~\cite{Monroe2005Impact,ahmad2017stability}:
\begin{equation}
    i = i_0^{ref}\exp\left[ \frac{\alpha\Delta G^{mech,eqbm}_{e^-}}{RT}\right] \left[  \exp\left\{ (1-\alpha)\frac{ F\eta}{RT} \right\}
-  \exp \left\{ -\alpha\frac{ F\eta}{RT} \right\} \right]
\end{equation}
Here, it is worth noting that $\Delta G^{mech,eqbm}_{e^-}$ is only used to denote the value of $\Delta G_M^{mech} - \Delta G_{M^+}^{mech}$. It does not refer to the real mechanical free energy of the electrons participating in the reaction under non-equilibrium conditions. Further, the overpotential $\eta$ in Eq.~\ref{eq:bv} includes contributions from the mechanical part of the free energy through the modified Nernst equation.
    
The values for $\delta^{mech}$, in general, may be material dependent and  require experimental investigation. The mechanical contribution to the free energies depends strongly on the molar volume tensor which also requires more experimental work similar to that of Pannikkat and Raj~\cite{pannikkat1999potential} to study the stress-dependent equilibrium potential. In this spirit, Carmona \etal~\cite{carmonaEffectMechanicalState2021} studied the effect of mechanical stress on the equilibrium potential at the Li metal-garnet electrolyte interface and found that the change is directly proportional to the applied stress and the molar volume of the metal. Further, the stress in the solid electrolyte had a negligible effect on the equilibrium potential. It has been argued based on the cation being part of the crystal lattice  of solid electrolyte and negligible swelling during ion transport in binary solid electrolytes that the molar volume of the Li ions in solid electrolytes should be zero~\cite{baraiMechanicalStressInduced2019,McMeeking2019metal}. However, there is no direct way to relate the volume change of the bulk solid electrolyte, which includes other species, to the molar volume of \ce{Li+}. Experiments need to be designed carefully on solid electrolytes with applied stress{/}pressure to maintain a conformal interface to separate the effects on potential from metal, the electrolyte, and the external forces.

\subsection{Constitutive Relations}
    
 The constitutive relations of solids relate the  strain to the stress developed and vice versa. In general, no single relation can be used for all values of the stresses and strains. For solids, there are generally three regimes that determine the relation: elastic, plastic, and creep. In the elastic regime, the deformation is microscopically caused by elongation or compression of bonds, resulting in a return to the pristine state before deformation when the load is released. The energy increases quadratically with  strain and the stress, being the derivative of energy with respect to strain, increases linearly with strain in this regime. For a crystalline solid, the stress tensor $\rtten{\sigma}$ is related to the strain tensor $\rtten{\epsilon}$ in the elastic regime by the relation:
\begin{equation}
    \rtten{\sigma} = \rtten{C} \rtten{\epsilon}
\end{equation}
where $\rtvec{C}$ is the elastic tensor of the crystal. For a polycrystal, one may perform an averaging over different crystal orientations which results in only two constants, the Young's modulus $E$ (GPa) and Poisson's ratio $\nu$:
\begin{equation}
    \rtten{\sigma} = \frac{E}{1+\nu}\rtten{\epsilon} + \frac{E\nu}{(1+\nu)(1-2\nu)}tr(\rtten{\epsilon})\rtten{I}
\end{equation}
where $\rtvec{I}$ is the identity tensor. The same relation can be used for polymers and amorphous solids like \ce{Li2S}-\ce{P2S5} solid electrolyte.

As the stress is increased beyond the yield strength of the material, it gives rise to slip, resulting in permanent plastic deformation of the material. The material undergoes strain hardening and the relation between uniaxial stress and strain is given by a power law:
\begin{equation}
    \sigma = \kappa \epsilon_p^n
\end{equation}
where $\epsilon_p$ is the plastic strain, $\kappa$ is the hardening modulus with units of stress, $n$ is the strain hardening exponent.

Note that the above two constitutive relations are independent of time. The creep constitutive relation, on the other hand, involves time dependence. During creep, the strain keeps increasing with time even at a constant stress value below the yield strength, resulting in permanent plastic deformation. The time rate of change of uniaxial strain due to creep $\epsilon_{\text{creep}}$ is related to the applied stress $\sigma$ as~\cite{lepageLithiumMechanicsRoles2019a}
\begin{equation}
    \frac{d\epsilon_{\text{creep}}}{dt} = A_c\sigma^m \exp\left(-\frac{Q_c}{k_B T}\right)
\end{equation}
where $A_c$ is a creep parameter dependent on the material and its microstructure, $Q_c$ is the activation energy for dislocation climb, $k_B$ is the Boltzmann constant and $T$  is the temperature. There are several mechanisms of creep such as dislocation creep, Nabarro-Herring creep, Coble creep, and Harper-Dorn creep.

    We now discuss the values of the material properties in the constitutive relations for Li metal and solid electrolytes. The elastic properties of Li have been measured through a variety of experiments and atomistic calculations. Using nanoindentation, Herbert \etal~\cite{herbertNanoindentationHighpurityVapor2018b} measured the value of elastic modulus of vapor-deposited polycrystalline Li to be $9.8\pm$11.9\% for 5 $\mu$m thick film and $8.2 \pm 14.5$\% for 18 $\mu$m thick film; similar techniques demonstrated that, at small length scales (indentention depths of 40nm), Li can withstand pressures that are up to 350 times larger than its policrystalline bulk yield stress~\cite{herbertNanoindentationHighpurityVapor2018}. Based on these values, the film surface is expected to be composed of mainly \hkl(111) and \hkl(100) orientations.
    Masias \etal~\cite{masiasElasticPlasticCreep2019} obtained a value of 7.82 GPa for the elastic modulus and 0.38 for the Poisson's ratio. The differences in the values agree with Xu \etal's~\cite{xuLi2017} calculations of the high degree of directional dependence, causing the elastic modulus to vary from 3 to 21.2 GPa. Further, the temperature dependence of the elastic properties of Li is rather weak~\cite{Slotwinski1969temperature,xuLi2017}.

    A variety of measurements exist for the value of yield strength of Li. The values range from 0.41 to 0.89 MPa~\cite{tariq2003li,schultz2002lithium,masiasElasticPlasticCreep2019} for bulk Li to as high as 105 MPa~\cite{xuLi2017} for Li micropillars. The large enhancement in yield strength  for smaller diameter micropillars is attributed to the size effect~\cite{Greer2008comparing,MinHan2013critical}. The hardening modulus and exponent of bulk Li are 1.9 MPa and 0.4 respectively. ~\cite{baraiLithiumDendriteGrowth2017,tariq2003li,schultz2002lithium}. Narayan and Anand~\cite{narayanLargeDeformationElastic2018} have developed a large deformation elastic-viscoplastic model for Li using the data from a nanoindentation study by Wang and Cheng~\cite{wangNanoindentationStudyViscoplastic2017}.

Strain rate has been found to considerably influence the mechanical stresses developed in Li metal~\cite{lepageLithiumMechanicsRoles2019a,masiasElasticPlasticCreep2019}. The low melting point of Li (453.5 K) combined with the low barriers for self-diffusion~\cite{Messer1975, Hao2018mesoscale,Jckle2018self} makes it very susceptible to creep.  For Li metal, the creep parameters have been found to be $m\sim 6.6 \pm 0.7$, $Q_c \sim 37 \pm 6$ kJ/mol and $A_c^{-1/m}\sim (3.0 \pm 0.55)\times 10^5$ Pa/s. The  creep has been explained through the Nabarro-Herring and Harper-Dorn mechanisms with diffusion and dislocation-mediated flow~\cite{herbertNanoindentationHighpurityVapor2018,masiasElasticPlasticCreep2019}. The mean pressure that Li can support was found to be $\sim$46-350 times higher than its yield stress~\cite{herbertNanoindentationHighpurityVapor2018}. 
Another study also achieved Coble creep in Li metal using a mixed electronic ionic conductor tubular matrix, which was beneficial for stabilizing Li metal electrodeposition~\cite{chenLiMetalDeposition2020}.

\begin{figure}
        \centering
        \includegraphics[scale=0.5]{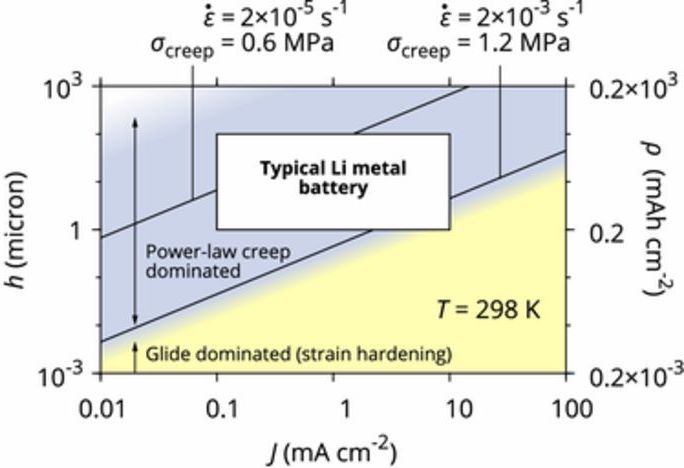}
        \caption{Dominant mechanics regimes for Li metal deformation under different current densities $J$ and Li anode thickness $h$. Li mechanics is shown governed by creep under strains encountered during Li-ion battery operation. During fast charging or the initial stage of plating in anode-free batteries, strain hardening may dominate. Reproduced with permission from Ref.~\cite{lepageLithiumMechanicsRoles2019a}.}
        \label{fig:mechanicsLi}
    \end{figure}

The strain rate $\dot{\epsilon}$ during Li plating under rigid boundary conditions can be related to the current density $J$ and Li anode thickness $h$ using the relation~\cite{lepageLithiumMechanicsRoles2019a}:
\begin{equation}
    \dot{\epsilon} = \frac{JV_{Li}}{hF}
\end{equation}
where $V_{Li}$ is the molar volume of Li metal and $F$ is the Faraday's constant. This relation can be used to obtain the dominant mechanics regime in Li. LePage \etal~\cite{lepageLithiumMechanicsRoles2019a} showed that under strains encountered during Li-ion battery operation, the relevant Li mechanics is creep-dominated, as shown in Fig.~\ref{fig:mechanicsLi}. Strain hardening is the dominant regime only during fast charging for very thin Li encountered in, for example, the initial plating stage of anode-free batteries.
    
Experiments and calculations of elastic properties of solid electrolytes abound in the literature. The full elastic tensor and moduli of different classes of solid electrolytes have been obtained from first-principles calculations~\cite{Deng01012016,Ahmad16uncertainty,yuElasticPropertiesSolid2016}. The elastic constants of ceramic solid electrolytes can be correlated to their brittleness through Pugh's ratio given as the ratio of shear and bulk modulus~\cite{Pugh1954}. In general, thiophosphates are the least brittle while antiperovskites are the most brittle among the different classes of solid electrolytes. The shear modulus of the garnet solid electrolyte \ce{Li7La3Zr2O12} (LLZO) and its Al/Ta doped analogues have been determined from acoustic and nanoindentation experiments to be  approximately 60 GPa~\cite{yuElasticPropertiesSolid2016}. The precise tuning of mechanical properties of solid electrolytes through processing techniques has also been demonstrated. The Young's modulus of \ce{Li2S}-\ce{P2S5} solid electrolyte can be doubled by densification through the application of an optimal molding pressure which reduces the porosity while maintaining the amorphous nature~\cite{garcia-mendezCorrelatingMacroAtomic2020}.
    
\subsection{Stability Analysis}
A linear stability analysis is useful to determine the initiation of uneven electrodeposition instead of the growth and propagation that follows. Linear stability analyses have been used widely in many other areas originating from solidification~\cite{MullinsSekerka63,MullinsSekerka64}, surface diffusion during cracking~\cite{Asaro1972} and fluid flows~\cite{THEOFILIS2004}. This has inspired many works on linear stability during metal electrodeposition~\cite{Pritzker1992,Elezgaray1998,Monroe2005Impact, Tikekar2016, ahmad2017stability, ahmad2017-anisotropy, khooLinearStabilityAnalysis2019}. 
Most of these models use linear elastic mechanical properties for the electrode and electrolyte, which is valid only for the initial response. However, models that describe the behavior at larger strain values need to incorporate plasticity and creep.

\begin{figure}
    \centering
    \includegraphics[scale=1]{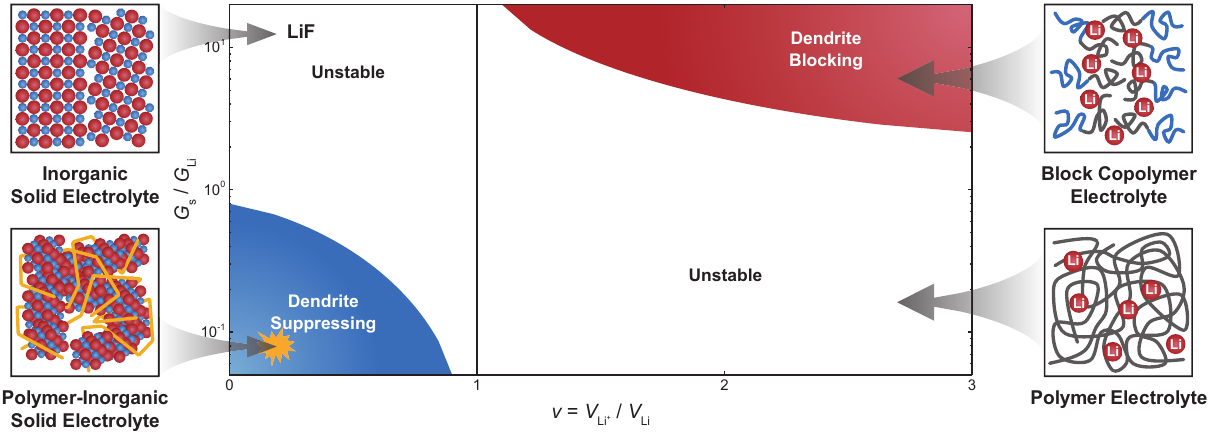}
    \caption{Stability diagram for electrodeposition at a solid-solid interface based on the shear modulus $G_s$ and $G_{Li}$ and molar volumes $V_{Li^+}$ and $V_{Li}$ for electrolyte and Li respectively. The stability regime and mechanism changes as the molar volume ratio $v=V_{Li^+}/V_{Li}$ crosses 1. Reproduced with permission from Ref. ~\cite{Fu2020universal}.}
    \label{fig:stabdiag}
\end{figure}

Monroe and Newman~\cite{Monroe2004Effect,Monroe2005Impact} developed the framework for a stability treatment of metal electrodeposition based on the modified Butler-Volmer kinetics and linear elasticity due to mechanical stresses. By using the properties of a polymer electrolyte in a Li-ion battery, they obtained the well-known stability criteria, requiring the shear modulus of the electrolyte to be approximately twice that of Li metal. Many studies have used the Monroe \& Newman criteria to materials beyond the class of polymers for which it was derived, such as inorganic solid electrolytes where the much smaller molar volume of Li flips the role of the hydrostatic and deviatoric stresses~\cite{ahmad2017stability}. With inorganic crystalline electrolytes, as the shear modulus increases, the hydrostatic stresses become destabilizing and the deviatoric, stabilizing. Stable electrodeposition is achieved when the shear modulus of the electrolyte is below a threshold value dependent on the molar volume ratio  of the metal ion in the electrolyte to the electrode. Fig.~\ref{fig:stabdiag} shows the two regimes for stability: dendrite blocking or pressure-driven stability and dendrite suppressing or density-driven stability, as well as the possible materials with the shear modulus and molar volume ratio for each regime~\cite{ahmad2017stability,Fu2020universal}. A composite electrolyte containing a polymer of intrinsic microporosity host and LiF guest was found to lie in the dendrite suppressing region, enabling an extension of cell life~\cite{Fu2020universal}. The importance of molar volume mismatch at the interface was further highlighted recently by Mistry and Mukherjee~\cite{mistryMolarVolumeMismatch2020}, who modeled the reaction and transport effects as well within solid electrolytes. \ce{Li3PS4}-like electrolytes were found to be more stable compared to garnet solid electrolytes with high shear modulus. Barai \etal~\cite{baraiEffectInitialState2017} used the condition for stress relaxation of freshly deposited Li at the interface in their model and found the current density dependence of the stability. At low current densities, they found that electrodeposition was always stable. 
Tikekar \etal~\cite{Tikekar2016} assumed fast reaction kinetics governed by diffusion and migration at the interfaces instead of Butler-Volmer equation and studied the effect of immobilizing a fraction of the anions on the electrodeposition. They found that using a combination of high stiffness polymer and immobilized anions can better suppress dendrite growth. McMeeking \etal~\cite{McMeeking2019metal} modified the boundary conditions used in the earlier linear stability analyses~\cite{Monroe2005Impact,ahmad2017stability} to satisfy force equilibrium without external traction and found that instability always exists at high current densities or long-wavelength perturbations at the interface regardless of the electrolyte stiffness. They, however, used a different choice of $\delta^{mech}$ compared to the earlier linear stability analyses. A systematic study of the effect of different parameters on stability including ionic conductivity was done by Tu \etal~\cite{tuElectrodepositionMechanicalStability2020} who found that variations in current density at the interface due to inhomogeneity in factors such as interfacial impedance, defects, solid-solid contact etc. can be reduced by decreasing the exchange current density and increasing the ionic conductivity. The linear stability analysis by Herbert \etal~\cite{herbertMechanismsStressRelaxation2019} aimed at connecting the changes in preferential Li electrodeposition to the stress intensification. This study identified a defect length scale at which material fracture can be caused due to the absence of stress relaxation mechanisms.
    
Considering the aforementioned importance of the molar volume in modulating the influence of stresses, we emphasize again the need for accurate measurements and calculations of the molar volume of Li in solid electrolytes in quantifying the electrochemomechanical effects. We note that the stability regime for inorganic solid electrolytes in Fig.~\ref{fig:stabdiag} does not change as long as the molar volume ratio remains below 1. It is important to identify the limits of the stability criteria based on the key phenomena modeled in stability analysis. SSBs have multiple failure mechanisms for different materials and under different operating and manufacturing conditions, hence a single stability analysis is usually not sufficient. Furthermore, future stability analyses can benefit from incorporating size effects and constitutive relations for real materials beyond the regime of elasticity.

\subsection{Pressure and Contact Effects}
Recently pressure has been identified as an important external tunable parameter to control the performance of SSBs. In some studies, high stack pressures have been found to benefit the cycling performance of the SSB. Krauskopf \etal~\cite{krauskopfFundamentalUnderstandingLithium2019} found that the interfacial resistance monotonically decreases with stack pressure for symmetric cells with LLZO solid electrolyte. Their work outlines that stack pressure can prevent void formation and contact loss at the interface with solid electrolyte due to its ability to cause Li plastic flow. This allows higher operation at current densities than the limit imposed by vacancy diffusion in Li. 
Li plastic flow was identified as a key factor responsible for reducing non-uniformities at the surface and improving interfacial contact through finite element modeling~\cite{tuElectrodepositionMechanicalStability2020}. However, a very high stack pressure may cause mechanical fracture by concentrating the stresses at defect tips, resulting in the existence of a mechanical stability window of pressure.
    
Zhang \etal~\cite{zhangRethinkingHowExternal2019} study the interface between a Li metal anode and the separator using a contact mechanics model. They find that even at high pressures, the contact between the electrode and separator is not conformal. The application of pressure results in a decrease of noncontact area and development of stresses in the Li anode which slows down electrodeposition at the asperities. A similar study with a solid  electrolyte~\cite{zhangPressureDrivenInterfaceEvolution2020} identifies the role of elastic, plastic, and creep stresses and roughness on the contact area. The study found that for a given stack pressure, a combination of plastic stresses and high roughness reduce contact area and hence, are undesirable. Elastic contact stresses at the interface, on the other hand, are beneficial, since elastic stresses are usually smaller and require a higher contact area for the same stack pressure.  Creep in Li also has the beneficial effect of increasing the contact area over time. The contact area scales linearly with the applied stack pressure.
    
An important distinction should be made between two types of stack pressure: fabrication pressure and operating pressure. For sulfide solid electrolytes, which typically do not require high temperature sintering, it has been found that a high fabrication pressure decreases the porosity of the solid electrolyte, resulting in enhanced cell performance and cycle life~\cite{douxPressureEffectsSulfide2020}. Operating stack pressure is essential to ensure optimal  contact at the interface and, thereby, affects the apparent ionic conductivity measured in experiments. Doux \etal~\cite{douxStackPressureConsiderations2020} found that only a modest operating stack pressure of 5 MPa is sufficient and high pressures may be detrimental and cause Li to creep through the pores of the electrolyte, resulting in shorting. This is at odds with the modeling work by Zhang \etal~\cite{zhangPressureDrivenInterfaceEvolution2020} which suggests a stack pressure of at least 20 MPa for conformal contact. We note that the stack pressure requirement varies depending not only on the class of solid electrolyte but also its microstructure, which needs to be accounted for in the modeling studies.
    
A key result from studies on stack pressure has been the distinction between its two effects in SSBs: the modification of Butler-Volmer kinetics at the contact regions and the ability of stack pressure to reduce noncontact regions, such as voids~\cite{kazyakLiPenetrationCeramic2020}. For the usual range of stack pressures $\sim$10 MPa, the second effect dominates. The first effect is considerably smaller due to the low overpotentials $\sim V_{\text{Li}} \sigma /F$ associated with stress $\sigma$~\cite{vermaMicrostructurePressureDrivenElectrodeposition2021a}. Larger stresses of the order $\sim$ 1 GPa have been found to develop within cracks in solid electrolytes where they may have a more pronounced effect on the kinetics~\cite{klinsmannDendriticCrackingSolid2019}.
The improvement of electrode/electrolyte contact and prevention of void formation is the major benefit of the application of stack pressure.

\begin{figure}[htbp]
    \centering
    \includegraphics[scale=0.7]{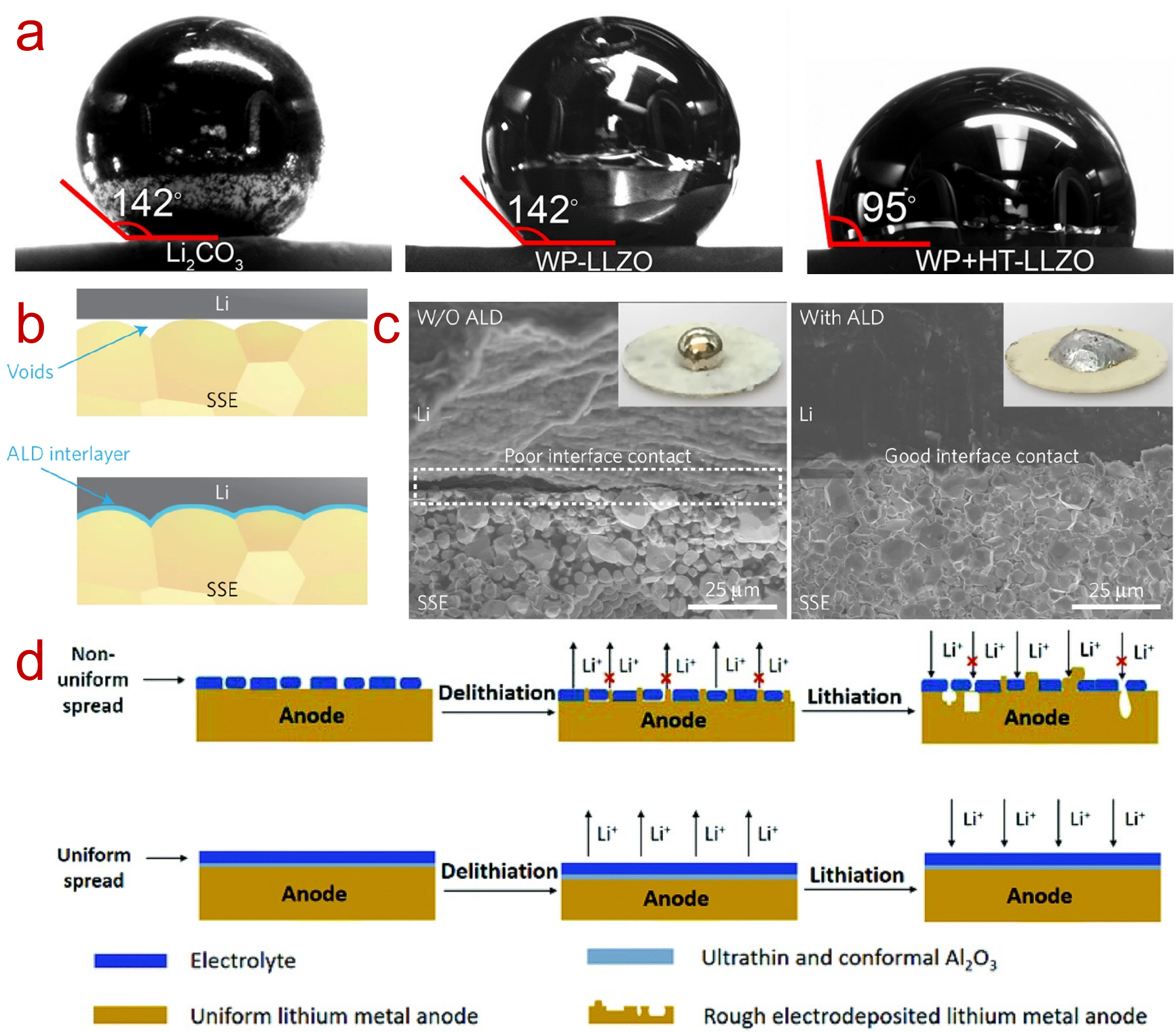}
    \caption{Wetting at the solid electrolyte/Li metal anode interface in a SSB. \textbf{(a)} Measured contact angle between molten Li metal and \ce{Li2CO3} surface, LLZO surface obtained on wet polishing, and LLZO surface obtained on  wet polishing and heat treatment at 500 $^{\circ}$C. Wet polishing and heat treatment clearly improves LLZO contact with molten Li metal. \textbf{(b)} Wetting of Li metal with the solid electrolyte enabled by \ce{Al2O3} ALD coating. \textbf{(c)}  Scanning electron microscopy images comparing the interfacial contact between Li metal and garnet solid electrolyte.  \textbf{(d)} Comparison of electrodeposition morphology between poorly wetted bare Li metal anode surface (top) and ALD \ce{Al2O3} coated Li metal anode surface (bottom). The ALD coated Li metal anode exhibits uniform Li metal electrodeposition by removing the voids present at the interface.  Panel \textbf{(a)} reproduced with permission from Ref. ~\cite{sharafiSurfaceChemistryMechanism2017}, \textbf{(b)}, \textbf{(c)} from Ref.~\cite{hanNegatingInterfacialImpedance2017}, and \textbf{(d)} from Ref. \cite{Chen2017lithium}.}
    \label{fig:wetting}
\end{figure}

A low contact angle between the Li metal anode and the solid electrolyte provides better wettability and is essential to reduce the interfacial resistance~\cite{wangCorrelatingInterfaceResistance2018}. Typical contaminants like \ce{Li2CO3} and \ce{LiOH} are detrimental to the wettability and have been known to increase the interfacial resistance~\cite{sharafiSurfaceChemistryMechanism2017}. Wet polishing and heat treatment of LLZO improves the wetting with Li metal as shown in Fig.~\ref{fig:wetting}a. Significant research has been devoted to the search for materials that can serve as coating layers to improve the wettability of the solid electrolyte with Li metal anode. 
An atomic layer deposited (ALD) \ce{Al2O3} coating between the Li metal anode and a  garnet solid electrolyte  drastically reduced the interfacial impedance~\cite{hanNegatingInterfacialImpedance2017, Chen2017lithium}. The authors found that the coating improved the wetting and ion transport at the Li metal/garnet solid electrolyte interface (Fig.~\ref{fig:wetting}b, c and d). 
Melt infusion of Li onto substrates decorated with lithiophilic additives such as strongly oxidizing oxides and organic coatings  has been used quite successfully to mitigate dendrite growth and large volume expansion with Li metal anodes~\cite{Zheng2020recent, Yu2018nanoflake, Wang2019tuning}.  Surface ozonolysis and ammoniation methods have also been applied to incorporate molten Li into carbon scaffolds to make composite anodes with improved cycle life~\cite{Tao2020surface}.

\subsection{Critical Current Density}
SSBs are characterized by a critical current density of plating and stripping beyond which long-term cycling cannot be achieved due to a variety of failure mechanisms. Extreme fast charging and discharge applications require that the battery be cycled at very high current densities, therefore, enhancing the critical current density of SSBs is an important goal.

Dendrite growth and propagation are the primary cause of the critical plating current density. The critical plating current density for LLZO has been found to lie in the range 0.05-0.2 mA/cm$^2$~\cite{Sharafi2016characterizing,Cheng2015inter,ishiguroStabilityNbDopedCubic2013}. Note that the actual microscopic critical current density may be higher than the nominal current density due to current focusing at favorable locations on the interface. The critical plating current density  has also been found to increase with temperature~\cite{sharafiCharacterizingLiLi7La3Zr2O122016} accompanied by a decrease in the interfacial resistance which remains upon cooling. However, another theory which put forward the non-zero electronic conductivity of solid electrolytes as the cause of dendrite growth~\cite{Han2019high} suggests that higher temperatures may not necessarily be beneficial. Another method to enhance critical plating current density is by improving surface microstructure resulting in reduced interfacial resistance and electrolyte densification~\cite{Cheng2015effect,sharafiCharacterizingLiLi7La3Zr2O122016,sharafiControllingCorrelatingEffect2017}. A theory accounting for coupling between electrochemistry and mechanics at the space charge layer at the Li metal/solid electrolyte interface also proposes raising the permittivity or increasing the surface energy of the Li-electrolyte interface to increase the critical current density~\cite{liDendriteNucleationLithiumconductive2019,Braun2015}.
    
The stripping current density has been found to be lower than the plating current density in SSBs and is the bottleneck for fast discharge applications. A critical stripping current density value of 0.1 mA/cm$^2$ has been found for the Li/LLZO solid electrolyte interface based on the self-diffusion rate in Li metal~\cite{kasemchainanCriticalStrippingCurrent2019}. The critical stripping current density can be increased by the application of stack pressure and alloying Li with metals like Mg~\cite{krauskopfDiffusionLimitationLithium2019}. The critical stripping current density is also expected to increase with temperature according to the Arrhenius relationship due to the higher self-diffusion rate of Li at higher temperatures. There is a step change in the critical current density as the temperature is raised beyond the melting point of Li. This indicates that mechanical properties of Li related to creep and flow, which undergo a step change upon melting, also influence the critical current density~\cite{Kinzer2021operando}.


\section{Failure Mechanisms}

\subsection{Crack propagation \& Fracture}

\begin{figure}[htbp]
    \centering
    \includegraphics[scale=0.6]{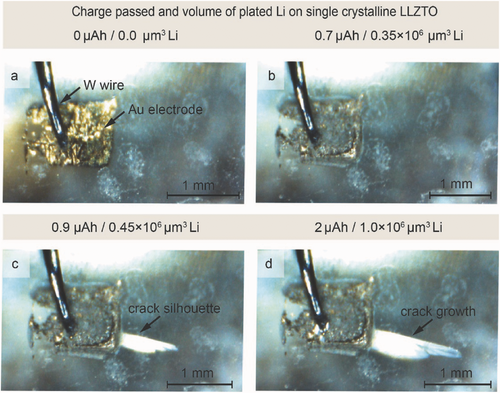}
    \caption{Crack propagation in single-crystalline  \ce{Li6La3ZrTaO12} solid electrolyte during galvanostatic Li deposition. The images \textbf{(a}-\textbf{d}) show successive snapshots as Li is deposited between the sputtered gold electrode and the solid electrolyte. Crack propagation is clearly seen in \textbf{c} and \textbf{d}. Reproduced with permission from Ref.~\cite{porzMechanismLithiumMetal2017}.} 
    \label{fig:crack}
\end{figure}

Two comprehensive studies on crack propagation in inorganic solid electrolytes were performed by Porz \etal~\cite{porzMechanismLithiumMetal2017} and Swamy \etal~\cite{swamyLithiumMetalPenetration2018}. Porz \etal~\cite{porzMechanismLithiumMetal2017} propose that Griffith flaws of size on the order of 1 $\mu$m and higher on the surface of solid electrolyte on which Li is plated  serve as hotspots for Li metal penetration. These flaws exist even in highly polished single crystalline solid electrolyte surfaces such as LLZO. The failure mechanism is the propagation of Li metal through cracks in the solid electrolyte as shown in Fig.~\ref{fig:crack} and focusing of stresses resulting in fracture according to the fracture criterion:
 \begin{equation}
     \sigma \geq \frac{K_{IC}}{\gamma \sqrt{\pi a}}
 \end{equation}
 where $K_{IC}$ is the fracture toughness of the solid electrolyte, $\gamma$ is a geometric factor and $a$ is the flaw length. 
 This criterion suggests soft electrolyte materials are prone to fracture.  With polycrystalline solid electrolytes, the Li propagates through the pore channels or grain boundaries (GBs). The results also show that failure is much easier during complete cycling rather than just plating. This points to a critical role of discharge in the failure as well. They find that even when the overpotential due to stress is a small fraction of the total overpotential ($\sim 1$\%), the stresses may become high enough to satisfy the criteria for brittle fracture of the solid electrolyte. However, in this study, the surface area between the electrode tip and the solid electrolyte available for electrodeposition is small, which may result in faster failure. The other study by Swamy \etal~\cite{swamyLithiumMetalPenetration2018} found that cracks may continue to grow without catastrophic failure even at short circuit. This study also points to the role of electrode boundaries in Li metal deposition where the current gets focused. In this study, the gold electrode was sputtered onto the solid electrolyte, which leaves little room to electrodeposit Li on it. Nevertheless, these studies provide a good understanding of the mechanisms of failure under severe conditions. These studies make the case for the use of amorphous solids as electrolytes instead of crystalline since the glassy \ce{Li2S}-\ce{P2S5} did not exhibit the same failure mode even up to a current density of 5 mA/cm$^2$.
 
 Modeling of crack propagation and fracture during electrodeposition is challenging and needs to incorporate the effects of stress intensification, electric field, the mechanical response of the solid electrolyte and Li, stochasticity, and imperfections. Bucci \etal~\cite{bucciModelingInternalMechanical2017a} studied intercalation-induced fracture during cycling in composite electrodes with solid electrolyte and found an upper limit on the volume expansion of the electrode particle and a lower limit on the fracture energy of the solid electrolyte necessary to prevent fracture. 
 Another set of modeling studies of cracks in solid electrolytes have found that a higher electrolyte conductivity, lower porosity, and higher fracture stress can enhance the critical current density for failure~\cite{rajCurrentLimitDiagrams2017,barroso-luqueAnalysisSolidStateElectrodepositionInduced2020}. It remains to be answered whether Li filled in cracks with high aspect ratios behaves as micro/nanoscale Li with much higher yield strength and mechanical anisotropy~\cite{xuEnhancedStrengthTemperature2017}. Li growing into cracks along specific orientations~\cite{ahmad2017-anisotropy,Cui2017-texturing} might present another challenge that theory needs to incorporate due to anisotropic mechanical properties. 
Klinsmann \etal~\cite{klinsmannDendriticCrackingSolid2019} report a swift rise in pressure of Li inside solid electrolyte cracks leading to crack propagation and short circuit. They use different boundary conditions compared to a similar study by Tu~\etal~\cite{tuElectrodepositionMechanicalStability2020}, namely that the Li in the crack cannot be pushed out and obtain a much higher pressure rise of $\sim$ 1 GPa. Stack pressure can limit the propagation of small-sized cracks by slowing down the reaction kinetics of Li growth inside the crack. However, even those cracks could eventually grow due to fatigue during cycling.
 
 Cracks in solid electrolytes typically propagate faster than Li can fill them during electrodeposition~\cite{ningVisualizingPlatinginducedCracking2021,swamyLithiumMetalPenetration2018}. Further, the nature of cracks is heavily influenced by the stability of the solid electrolyte interphase (SEI) formed between the solid electrolyte and the metal anode. With growing SEI, cracking may be caused by the strain due to volume expansion, while with stable SEI, cracking initiates through spallations at the solid electrolyte. Crack propagation studies have emphasized the role of boundaries in failure. The strains are the highest close to the edges of the electrode where cracking happens due to strain accumulation during cycling~\cite{ningVisualizingPlatinginducedCracking2021}.
A single solid electrolyte may produce a range of morphologies when Li metal penetrates leading to failure. Four different types of morphologies of penetrated Li, namely straight, branching, spalling and diffuse were observed with LLZO solid electrolyte~\cite{kazyakLiPenetrationCeramic2020}. While the driving force (current density) and local microstructure are key factors influencing the morphology, further investigations are required to decouple the influence of more fundamental properties such as mechanical properties of Li and geometry of the crack tip. 
Qi \etal~\cite{Qi2020anew} propose the use of compressive stress to eliminate cracks similar to what is done for glasses while ensuring the ionic conductivity of the electrolyte is not dramatically reduced. They find that LLZO solid electrolyte does not suffer a drastic loss in ionic conductivity due to compressive stress. This may not apply to all solid electrolytes. For example, the diffusivity of polyborane solid electrolytes is reduced by three orders of magnitude with a 5\% compressive strain~\cite{varley2017polyborane}. Ye and Li~\cite{Ye2021dynamic} design a multilayer interface between Li metal and solid electrolyte to dynamically suppress Li penetration in the cracks through the filling of cracks by localized decompositions. Such dynamical driving forces to suppress Li penetration are promising in the quest for fast charging and stable cycling SSBs since surface defects are very hard to eliminate during manufacturing and processing.

\subsection{Pitting}
Electric aircraft applications, especially vertical takeoff and landing aircraft, demand a high power discharge from their energy source~\cite{fredericksPerformanceMetricsRequired2018,epsteinConsiderationsReducingAviation2019,billsPerformanceMetricsRequired2020}. This requirement can cause a different, less studied, failure mechanism during the Li stripping process, in which voids and pits form on the interface between the anode and the electrolyte~\cite{wangCharacterizingLiSolidElectrolyteInterface2019,woodDendritesPitsUntangling2016,cohenMicromorphologicalStudiesLithium2000,shiLithiumMetalStripping2018}. The presence of voids and pits can significantly change the morphology of the anode surface,~\cite{lewisLinkingVoidInterphase2021,sanchezLithiumStrippingAnisotropic2021,kasemchainanCriticalStrippingCurrent2019} and, thus, have substantial effects on the subsequent deposition cycles~\cite{Monroe2005Impact,ahmad2017stability,kasemchainanCriticalStrippingCurrent2019}.

The pitting critical current density is lower than the plating one~\cite{wangCharacterizingLiSolidElectrolyteInterface2019,kasemchainanCriticalStrippingCurrent2019} but also with a greater dependence on the applied stack pressure~\cite{kasemchainanCriticalStrippingCurrent2019}. Therefore, one potential avenue for preventing this issue would be the application of stack pressure, which could raise the pitting critical current density by a factor of 5 from 3 to 7 MPa~\cite{kasemchainanCriticalStrippingCurrent2019}.
In order for more attainable mitigation strategies to be proposed, an understanding of the underlying process of void and pit formation is needed. 

\begin{figure}[htbp]
    \centering
    \includegraphics[width=0.9\textwidth]{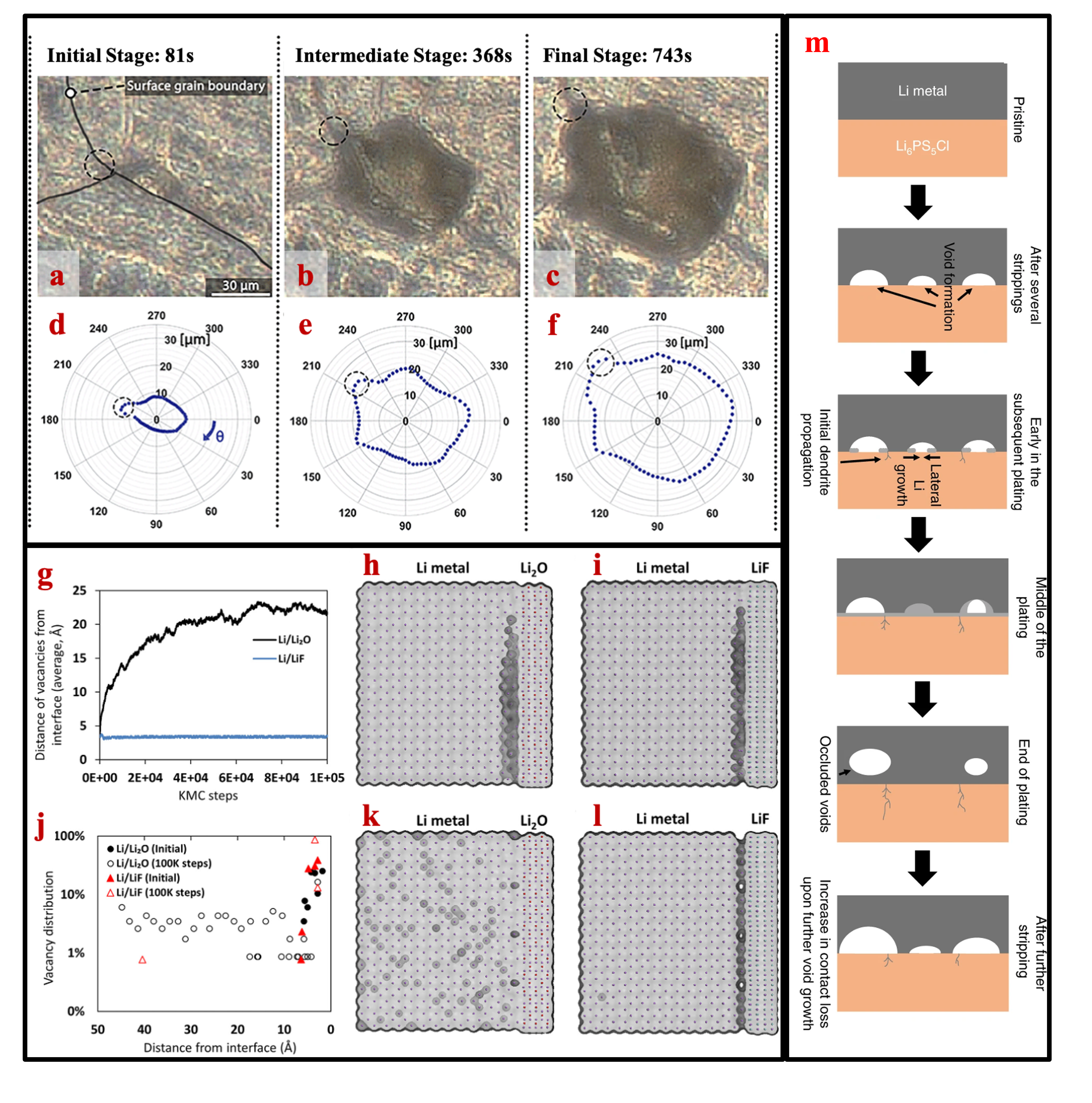}
    \caption{Images and depictions of the formation and evolution of pits, and migration of vacancies at and near interfaces of Li metal with other materials. \textbf{(a-c)} Optical images of the evolution of a pit, with surface GBs outlined in black in panel \textbf{(a)}. \textbf{(d-e)} Polar plots of the distance between the edges and the center of the pit at times 240s, 430s, and 730s, respectively. Panel \textbf{(g)} shows the average distance between vacancies and anode surface as a function of a number of kinetic Monte Carlo (KMC) steps taken. The initial configurations are displayed in panels \textbf{(h)} for Li/\ce{Li2O} and \textbf{(i)} for Li/LiF, while the configuration at $10^5$ KMC steps are represented in panels \textbf{(k)} for Li/\ce{Li2O} and \textbf{(l)} for Li/LiF. It is evident that vacancies in Li metal are much more likely to diffuse into bulk when \ce{Li2O} is present at the interface, instead of LiF. Panel \textbf{(m)} is a schematic of one of the proposed hypotheses for pit formation: as voids are formed during stripping and then occluded during plating, they can merge with other voids on subsequent stripping processes and create large pits. Panels \textbf{(a-f)} are reproduced with permission from Ref. \cite{sanchezLithiumStrippingAnisotropic2021}, \textbf{(g-l)} from Ref. \cite{yangMaintainingFlatLi2021}, and \textbf{(m)} from Ref. \cite{kasemchainanCriticalStrippingCurrent2019}.}
    \label{fig:pitting}
\end{figure}

Various hypotheses have been formulated to explain the origin of this pitting behavior. Fundamentally, stripping involves the transport of Li ions from the electrode to the solid electrolyte, resulting in Li vacancies in the electrode close to the interface, which then diffuse into the bulk. If the stripping process is too fast, multiple vacancies can be generated close to the interface without being replenished with Li atoms due to vacancy diffusion, resulting in the formation of voids. Krauskopf \etal~\cite{krauskopfFundamentalUnderstandingLithium2019} obtained a fundamental limit on the current density at the Li metal anode limited by vacancy diffusion equal to 0.1 mA/cm$^2$. This value is quite low to enable fast charging applications with planar Li metal electrodes. Some studies suggest that the electrode microstructure plays an important role in the growth of voids, and demonstrate that pit nucleation and growth happens preferentially along surface GBs~\cite{sanchezLithiumStrippingAnisotropic2021, sanchezPlanViewOperandoVideo2020}, as shown in Fig. \ref{fig:pitting}. Others suggest that some of the voids formed during stripping do not get filled in the plating process due to the lack of contact with the electrolyte~\cite{kasemchainanCriticalStrippingCurrent2019}, and, therefore, cause current to concentrate at the void edges. Besides aiding dendrite formation, this current focusing can occlude the void beneath the anode surface~\cite{kasemchainanCriticalStrippingCurrent2019}. During following stripping cycles, these occluded voids can merge with other voids (occluded or new) and form pits. This process is exemplified in Fig.~\ref{fig:pitting}. 

The studies mentioned above indicate that the most promising direction to tackle the pitting problem is to better comprehend the formation of voids, which are simply an agglomeration of Li vacancies at the anode surface. A first-principles density functional theory (DFT) study of Li surfaces indicates that the \hkl(100) and \hkl(110) facets are prone to suffer vacancy clustering, while the \hkl(111) facet should be more resistant to this issue~\cite{venturiThermodynamicsLithiumStripping2021}; however, the same work also shows that different interfaces can have a significant impact on the stripping behavior. Another theoretical work combining first-principles and Monte Carlo methods showed that clever interface engineering can also help in mitigating pit formation: when vacancies are formed at a Li surface, they have a higher tendency of migrating to bulk regions if the surface is coated with \ce{Li2O} than if it is coated by LiF~\cite{yangMaintainingFlatLi2021}. While these recent studies shed some light on the process of void formation, a deeper understanding of the overall pitting process, from both a nano and microscale perspective, is still needed if we desire mitigation routes for Li metal SSBs.

\subsection{Growth of dendrites through grain boundaries}

\textit{Chemomechanical aspects}. Although numerous strides have been made in understanding the formation and growth of dendrites at the anode-electrolyte interface, the effects that the solid electrolyte defects, such as GBs, can have on the Li deposition process remains somewhat elusive. One of the earliest observations of Li dendrite growth in garnet solid electrolytes by correlating them to the presence of GBs and pores was made by Imanshi and co-workers~\cite{ishiguroStabilityNbDopedCubic2013,sudoInterfaceBehaviorGarnettype2014}. Further direct evidence was obtained through scanning electron microscopy~\cite{renDirectObservationLithium2015}. Recently, many more experimental studies have shown that Li dendrites can infiltrate and grow along the GBs of garnet electrolyte  LLZO    ~\cite{Han2019high,chengIntergranularLiMetal2017,tsaiLi7La3Zr2O12InterfaceModification2016}. These results indicate a need for a better understanding of the micro and nanoscale processes that take place in GBs of solid electrolytes. 

\begin{figure}[htbp]
    \centering
    \includegraphics[width=\textwidth]{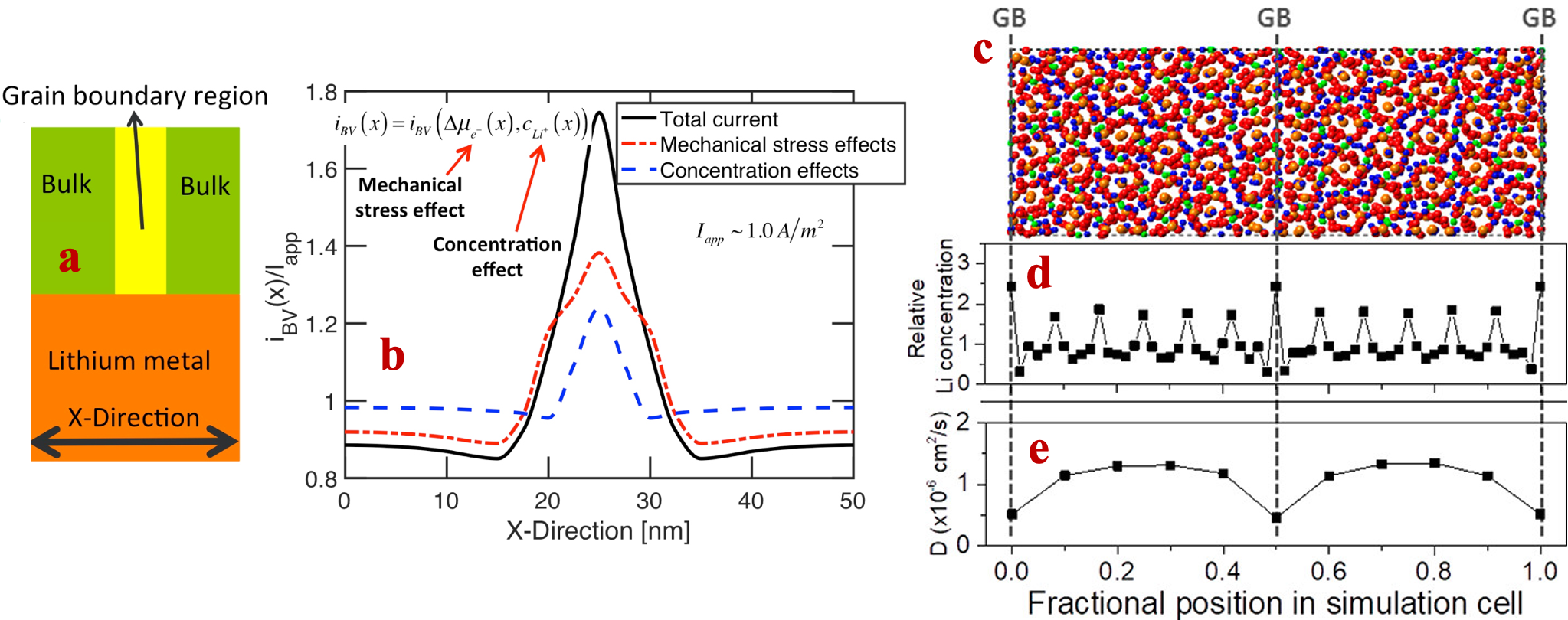}
    \caption{Computational studies of chemomechanics of solid electrolyte GBs. The interface between an LLZO GB and a Li surface is represented in \textbf{(a)}. The current focusing at the GB region is shown in panel \textbf{(b)}, with contributions from both the mechanical and the concentration effects depicted separately. Schematic of a $\Sigma3$ GB is illustrated in panel \textbf{(c)}. Relative concentration and diffusivity of Li atoms are plotted as a functions of the fractional position along the GB simulation cell in panels \textbf{(d)} and \textbf{(e)}, respectively. It is evident that these GBs lead to a higher concentration of Li atoms, while at the same time, reducing their mobility. Panels \textbf{(a)} and \textbf{(b)} are reproduced with permission from Ref. \cite{baraiRoleLocalInhomogeneities2020}, and panels \textbf{(c-e),} from Ref. \cite{yuGrainBoundarySoftening2018}.}
    \label{fig:gb_chemomec}
\end{figure}

The correlation between dendrite-blocking capability and elastic moduli makes it essential to quantify the mechanical response of GBs. In spite of being an experimental challenge, the calculation of the full stiffness tensor of GBs is a relatively straightforward process through \textit{in silico} simulations. Molecular dynamics (MD) based computational studies have already demonstrated that GBs in one of the most promising solid electrolyte candidates, LLZO, suffer from ``softening:'' the moduli in these regions can be up to 50\% smaller than in bulk,~\cite{yuGrainBoundarySoftening2018, baraiRoleLocalInhomogeneities2020} as can be seen in Fig. \ref{fig:gb_chemomec}.
    
According to the model developed by Monroe and Newman~\cite{Monroe2005Impact}, this softening can lead to current localization at the interface between GB and metal anode~\cite{baraiRoleLocalInhomogeneities2020}, as shown in Fig.~\ref{fig:gb_chemomec}. Such concentration of current is fundamental for the nucleation of dendrites at the interface between anode and solid electrolyte GB. After formation,  propagation of these dendrites through the GB network is nearly inevitable, and its growth velocity was shown to be positively correlated with the yield strength of Li at the nanoscale~\cite{baraiRoleLocalInhomogeneities2020}.

Another factor that can induce current focusing is a higher concentration of charge carriers at the GBs~\cite{baraiRoleLocalInhomogeneities2020}. Using similar MD techniques, it has been demonstrated that the concentration of Li atoms at the most common LLZO GBs is higher than in bulk by at least 50\%~\cite{yuGrainBoundaryContributions2017}, as represented Fig. \ref{fig:gb_chemomec}. Interestingly, the GBs studied also presented a higher oxygen composition than bulk LLZO, suggesting the possibility of formation of \ce{Li2O} phase between grains~\cite{yuGrainBoundaryContributions2017}. While such an enhanced concentration of Li atoms contributes to localization of the ionic current at the GBs, it is partially mitigated by the reduced diffusivity of Li in these regions, which can be as low as 33\% of the corresponding bulk values~\cite{yuGrainBoundaryContributions2017}.
    
The presence of GBs is intrinsically related to the formation of voids and pores in the solid electrolyte. These voids are particularly prone to Li penetration, since they offer little to no mechanical resistance to the growth of dendrites. The volume and concentration of these pores, as well as the overall void network, can be probed experimentally with techniques such as X-ray tomography. These measurements have been fundamental in showing that not only overall porosity level~\cite{dixitTortuosityEffectsGarnetType2019}, but also void interconnectivity,~\cite{shenEffectPoreConnectivity2018} are both intimately related with lowering the critical current density of solid electrolyte based Li batteries. Spallation cracks observed in solid electrolytes also grow fastest towards regions with high porosity~\cite{ningVisualizingPlatinginducedCracking2021}

It is also worth noting that several studies of solid electrolytes have been inspired by the success of thin film batteries based on vapor deposited Li phosphorus oxynitride (LiPON), which achieves stable cycling for thousands of cycles~\cite{liSolidElectrolyteKey2015,Bates1993,Bates2000}. However, their lack of a competitive edge against the conventional Li-ion batteries due to low rate capability and thin film architecture has rendered them useful mostly as model SSBs with stable cycling~\cite{albertusStatusChallengesEnabling2018}. The reason for the stable cycling of LiPON is thought to be its homogeneous GB-free surface. This was further corroborated by an experiment involving Li plating on copper current collector with a LiPON-LiPON interface~\cite{westoverDepositionConfinementLi2019}. The Li initially started plating on the current collector but then expanded towards the two dimensional plane of the LiPON-LiPON interface. The confinement of plated Li at this interface is an indication of the dendrite suppression by LiPON solid electrolyte. Along a similar theme, amorphous LLZO has shown promise over crystalline LLZO for stable cycling~\cite{sastreBlockingLithiumDendrite2021}. However, the high electronic conductivity of other solid electrolytes in comparison to LiPON is also being considered as a key reason for dendrite growth~\cite{Han2019high}.

One potential avenue for dealing with these GB and void issues lies in the use of additives, which have the potential to modify the structure of the interfaces between grains during the sintering process required to manufacture solid electrolytes. For instance, in the case of tantalum (Ta) doped LLZO (LLZTO), the introduction of a \ce{Li6Zr2O7} additive (LZO) was demonstrated to increase the dendrite-suppressing capabilities of the sintered electrolyte~\cite{zhengGrainBoundaryModification2021}. Since, during sintering, LZO decomposes into \ce{Li2O} and \ce{Li2ZrO3}, the role it plays in improving electrolyte performance is two-fold: \ce{Li2O} serves as a mother powder, alleviating \textit{in situ} Li losses, while \ce{Li2ZrO3} fills up the voids in the structure and modifies the composition of the GBs, strengthening intergranular bonding, enhancing the mechanical response of these regions, and therefore making them more resistant to dendrite penetration~\cite{zhengGrainBoundaryModification2021}. The use of additives, however, can come at a cost -- employing \ce{Al2O3} additive in LLZTO preparation can lead to the formation of GBs containing \ce{LiAlO2}, which have high mechanical strength, but low Li-ion conductivity~\cite{huangEnhancedPerformanceLi2020}.

This myriad of studies indicates that the impact solid electrolyte GBs can have on dendrite-free, stable Li electrodeposition on a metallic anode is significant. While abundant progress has been made in uncovering the extent of this effect, further research is still needed to tackle lingering concerns. In particular, a full discovery of all the phases that may be present at different GBs of a candidate electrolyte is needed. Additionally, insights into the chemomechanics of such potentially multi-phase regions can aid in engineering grain interfaces that are dendrite-resistant, ionic conducting, while also blocking electron leakage from the anode. 

\if
\vv{For this next part, it might be interesting to merge it with doping, since these additives have mostly been used in tantalum doped LLZO: LLZTO} Mention possibility to address some of these issues with LZO additive in LLZT: LZO and its decomposition products get trapped in LLZT GBs doing sintering and that somehow improves performance ~\cite{zhengGrainBoundaryModification2021} \\
Other additives might help with dendrite suppression but at a cost of ionic conductivity ~\cite{huangEnhancedPerformanceLi2020}\\ \cite{baraiMechanicalStressInduced2019}\\
This\cite{shenEffectPoreConnectivity2018,dixitTortuosityEffectsGarnetType2019}\\
\cite{neumannEffect3DStructure2021}\\
\vv{add transition to next sub-sub-section}
\fi
    
\textit{Electronic conductivity}. Besides facilitating dendrite formation and growth from a chemomechanical standpoint, solid electrolyte GBs can pose an additional concern to battery safety and operation: their electronic conductivity. Nucleation and growth of dendrites has been experimentally observed at random locations in the solid electrolyte, without being directly connected to the metallic anode~\cite{hanHighElectronicConductivity2019,aguesseInvestigatingDendriticGrowth2017,songProbingOriginElectronic2019}.

\begin{figure}[htbp]
\centering
\includegraphics[width=\textwidth]{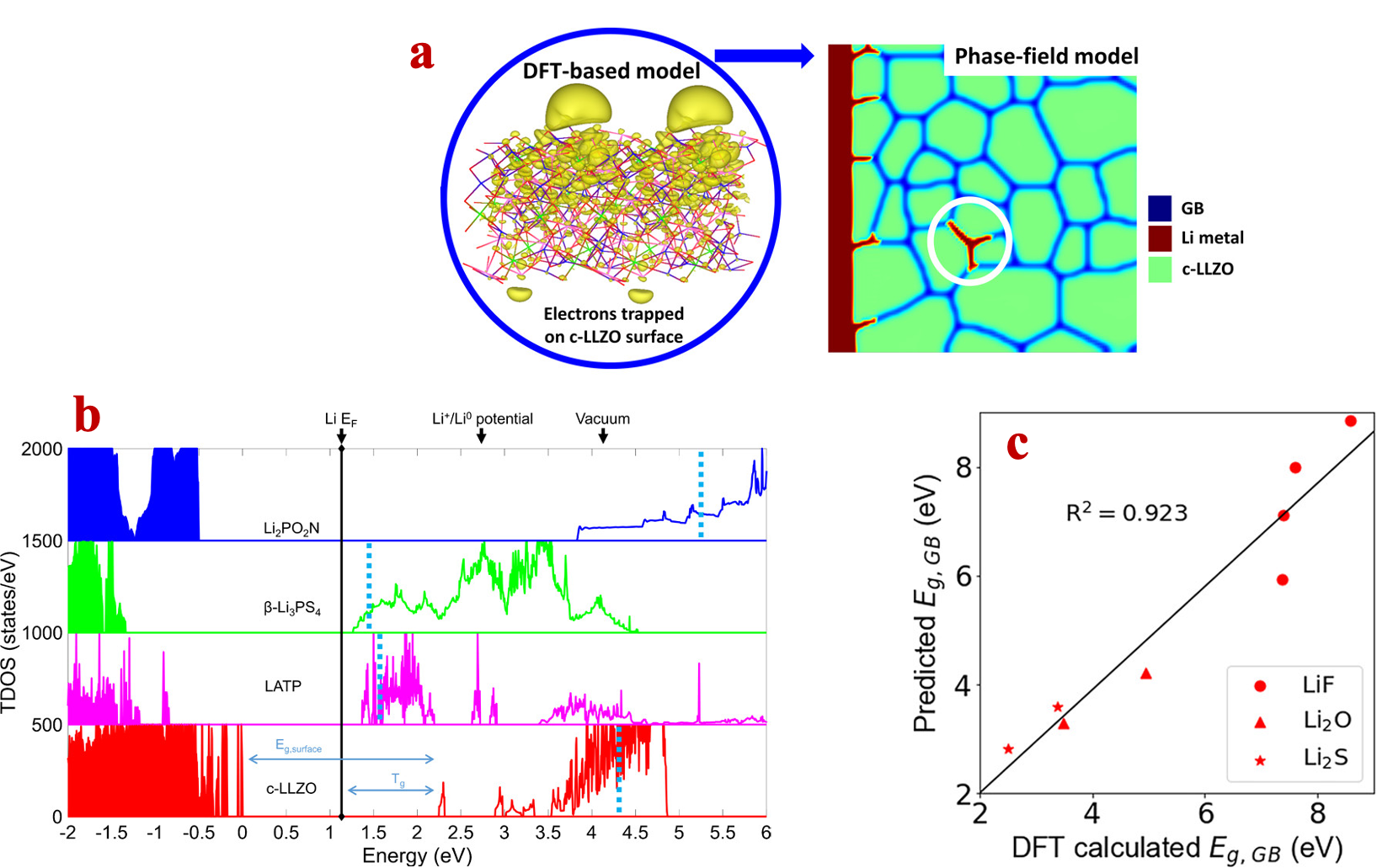}
\caption{Study of electronic properties of GBs. Panel \textbf{(a)} displays, on the left, a schematic of the charge isosurfaces of an LLZO surface with excess electrons, demonstrating their tendency to accumulate at the surface of this material. Phase field models that incorporate this insight can show the possibility of Li reduction inside the solid electrolyte GBs, as shown on the right image of panel \textbf{(a)}. The density of states of surfaces of four promising solid electrolytes is presented in panel \textbf{(b)}, where the dotted blue line represents the corresponding bulk conduction band minimum (CBM) level. It is clear that both LiPON and LLZO have surfaces with a significantly reduced band gap. In the case of LLZO, the Li reduction issue along GBs is exacerbated by the fact that the surface CBM is very close to the Li$^0$/Li$^+$ potential, thus facilitating the transfer of electrons from the anode to the electrolyte. The relationship used to calculate band gaps of GBs in Equation \ref{eq:GB_gap} is shown in panel \textbf{(c)}, along with data from the same work.~\cite{fengImpactElectronicProperties2021} Panels \textbf{(a)} and \textbf{(b)} are reproduced with permission from Ref. \cite{tianInterfacialElectronicProperties2019}, and panel \textbf{(c)}, from Ref. \cite{fengImpactElectronicProperties2021}.}
\label{fig:qiElectronicGB}
\end{figure}

The most likely explanation for this behavior is that electronic conduction is somehow facilitated at the GB regions. High-resolution transmission electron microscopy, coupled with electron energy loss spectroscopy experiments have determined that the electronic band gap of LLZO solid electrolyte is significantly reduced at the boundaries between grains: while the gap at the interior of the grains is close to 5.9 eV; at the GBs, it can be lower than 2 eV~\cite{liuLocalElectronicStructure2021}. Such a reduction in band gap makes GBs prime channels for electron leakage, allowing \ce{Li+} ions to prematurely reduce in the interior of the SE, instead of at the electrode. This process leads to the formation of isolated \ce{Li^0} filaments inside the SE, as shown in Fig. \ref{fig:qiElectronicGB}~\cite{tianInterfacialElectronicProperties2019}. These filaments can become connected upon further cycling and eventually result in a short circuit. 

This hypothesis of enhanced electronic conduction at the GBs is further corroborated by first-principles DFT studies that evaluated how the band gap of solid electrolytes changes from bulk to surface. In their work, Tian \etal~\cite{tianInterfacialElectronicProperties2019} show that cubic LLZO (c-LLZO) and LiPON suffer significant band gap reductions at their surfaces, while \ce{Li_{1.17}Al_{0.17}Ti_{1.83}(PO4)3} (LATP) and $\beta$-\ce{Li3PS4} surfaces maintain their bulk gap. This reduction in band gap can push the conduction bands below the $\text{Li}^{+}/\text{Li}^0$ Li plating potential, which could allow electrons to transfer from the metallic anode to the electrolyte~\cite{tianInterfacialElectronicProperties2019}. It was shown that, if such a transfer indeed occurs, these excess electrons would most likely localize at the surfaces of c-LLZO, but would remain beneath the surfaces of LiPON, LATP, and $\beta$-\ce{Li3PS4}. These results for c-LLZO were incorporated into phase-field simulations to demonstrate the viability of isolated dendrite nucleation in the GBs of the solid electrolyte, as shown in Fig.~\ref{fig:qiElectronicGB}.

It is worth mentioning that this reduction of band gap at GB regions has also been observed in other crystalline materials relevant for battery applications, such as LiF, \ce{Li2O}, and \ce{Li2S}, all of which have some GBs with a reduced band gap as well as at material interfaces~\cite{Ahmad2021interfaces}. Feng \etal~\cite{fengImpactElectronicProperties2021} propose that the band gap at the GB can be calculated using the relation
\begin{equation}
\label{eq:GB_gap} 
E_{\text{g,GB}} = \frac{d_{hkl}}{W_{\text{GB}}}\left[E_{\text{g,bulk}} - E_{\text{g,surf}}\right] + E_{\text{g,surf}},
\end{equation}
where $E_{\text{g,GB}}$, $E_{\text{g,bulk}}$, and $E_{\text{g,surf}}$ correspond to the band gaps at the GB, bulk, and surface, respectively; $d_{hkl}$ is the distance between the \hkl(hkl) Miller planes in the bulk structure, and $W_{\text{GB}}$ is the width of the GB, defined as the maximum distance between the \hkl(hkl) planes at the GB region that contain undercoordinated atoms. This relationship was shown to adequately capture the band gap at GBs, as represented in Fig.~\ref{fig:qiElectronicGB}, and indicates that, the further there exist uncoordinated atoms from the grain interface, the less the GB band gap behaves as bulk, and the more similar it is to that of a free surface. This result is one of the first to suggest a GB engineering design rule, capable of circumventing computationally expensive first-principles calculations. 

Broadly, a formal theory of electronic conductivity in GBs is still deficient. Even in the cases where a reduced band gap has been observed, it generally remains significant, usually at values greater than 3 eV. Due to the computational challenges associated with simulating interfaces, data for further investigations is also lacking. The same applies to interfaces between grains of different crystals: while several valuable observations can be obtained from nano and microscacle modeling, such as space-charge analysis, understanding the atomic level behavior at these regions can provide useful insight for better comprehending both ionic and electronic mobility inside battery materials.

\section{Manufacturing, Processing, and Pack Integration}
\begin{figure}[htbp]
    \centering
    \includegraphics[width=\textwidth]{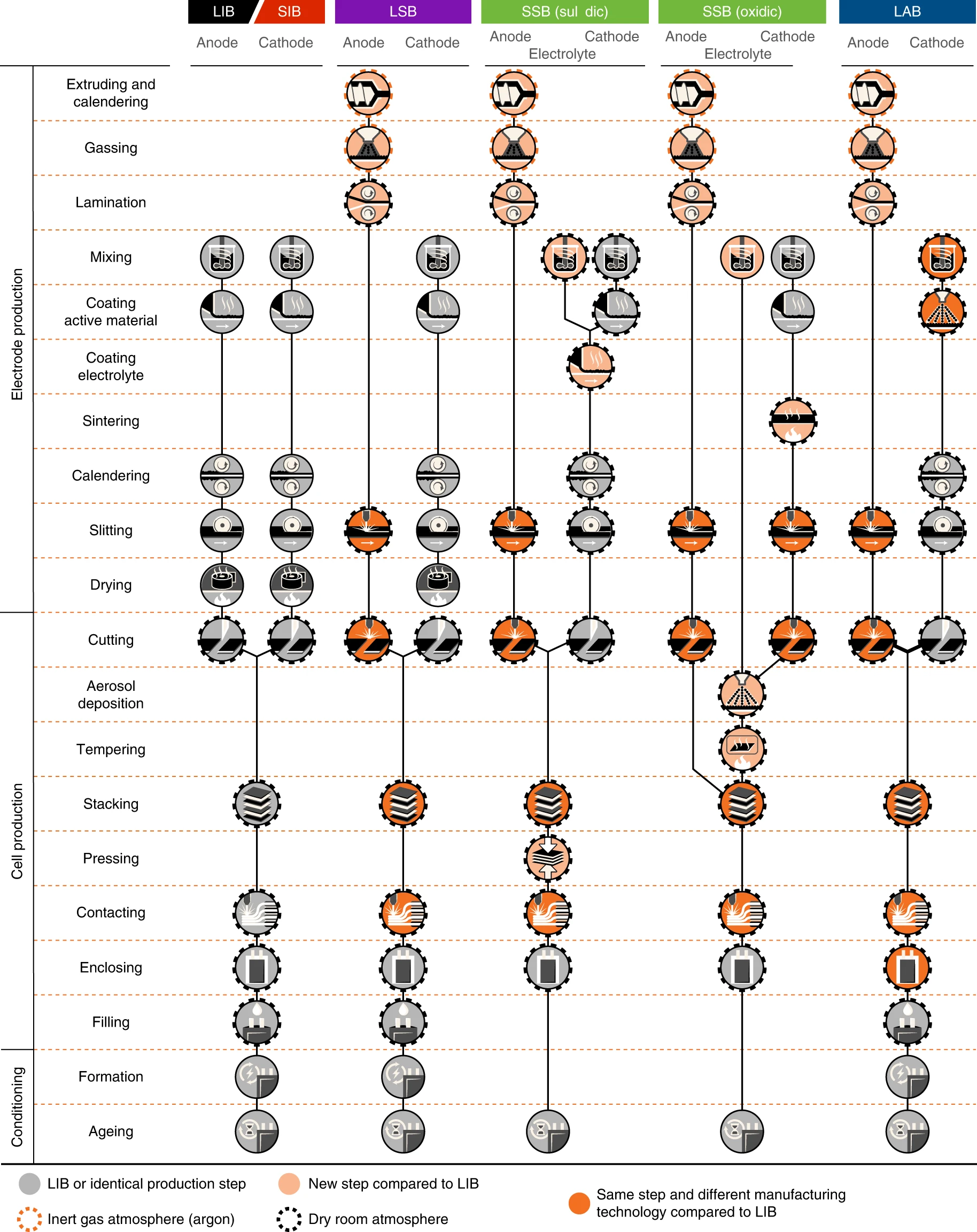}
    \caption{A comparison of the steps to manufacture Li-ion (LIB), Na-ion (SIB), Li-sulfur (LSB), SSB, and Li-air (LAB) batteries. Reproduced with permission from Ref.~\cite{duffner2021post}.}
    \label{fig:manufacturing}
\end{figure}

Manufacturing solid electrolytes and incorporating them within a cell is arguably very different from traditional liquid electrolyte based batteries~\citep{hatzell2021prospects,huang2021manufacturing,duffner2021post,wang2020enabling, wangTransitioningSolidstateBatteries2021}. Solid electrolytes generally perform a dual function of conduction of ions and that of a solid electrode separator/electronic insulator. A step-by-step comparison of the difference in the manufacturing process for a conventional Li-ion battery with a liquid electrolyte and SSBs with sulfide-based electrolyte and oxide-based electrolytes, compiled by Duffner \etal~\cite{duffner2021post} is shown in Fig.~\ref{fig:manufacturing}. Several different steps in manufacturing of SSBs can have implications on the mechanical properties of the electrolyte. For instance, one of the main differences between sulfidic and oxidic solid electrolytes is that sulfidic electrolytes, unlike their oxidic counterparts, are not stable against Li metal, and thus require a protective layer to protect the electrode-electrolyte interface~\cite{duffner2021post}. For sulfidic solid electrolytes, the protective layer determines the interfacial impedance and the stability of Li plating and stripping. The requirement for formation cycling changes the pressure regimes under which the cell stacks can be assembled. This leads to differences in grain size, density, and pores, giving rise to different chemomechanical regimes of operation for the SSB.

Schnell \etal~\cite{schnell2018all,schnell2019prospects} discuss the various processing steps possible for the cathode and solid electrolyte assembly. Two popular approaches discussed are (i) a cathode composite layer which includes about 30\% by volume of solid electrolyte, (ii) a separate solid electrolyte matrix that is stacked along with the cathode~\cite{schnell2018all,placke2017lithium}. The former leads to regimes where the ionic conductivity can be correlated to the slurry homogeneity~\cite{duffner2021post}. For sulfidic solid electrolytes, Duffner \etal~\cite{duffner2021post} note that the cathode composite requires an additional layer of pure solid electrolyte to complete the cathode-solid electrolyte sub-assembly. This kind of sub-assembly has unique requirements in terms of stack pressure which affects the mechanical behavior of the solid electrolyte and Li metal through the compressive stress~\citep{hatzell2021prospects}.

Wang \etal~\cite{wang2020enabling,wang2020electrochemical} have evaluated the manufacturing and processing aspects related to the anode side of the SSB. They specifically evaluate the feasibility of manufacturing \textit{Li-free} cell assemblies where the Li anode is formed at the interface of the anode current collector and the solid electrolyte while Li is supplied by the lithiated cathode material. This is quite similar in spirit to the experimental setup by Porz \etal~\cite{porzMechanismLithiumMetal2017}. One of the main aspects of investigation is the nucleation and growth of the Li anode at the solid electrolyte and current collector interface. They propose three mechanisms for nucleation and growth, (i) growth leading to separation of the current collector and solid electrolyte, (ii) horizontal plastic flow of Li, (iii) deposition at the nucleate center~\cite{wang2020enabling}. Each of these mechanisms is governed by the chemomechanical properties of materials at the interface. Another important aspect that affects the processing conditions, as well as the mechanical properties of the solid electrolyte, is the thickness of the electrolyte layer itself~\cite{balaish2021processing}. Thinner electrolytes provide avenues to reduce cost as well as exploit cell energy density further. However, they may undergo failure due to cracks and shorting much quicker. 

\begin{figure}[htbp]
    \centering
    \includegraphics[width=\textwidth]{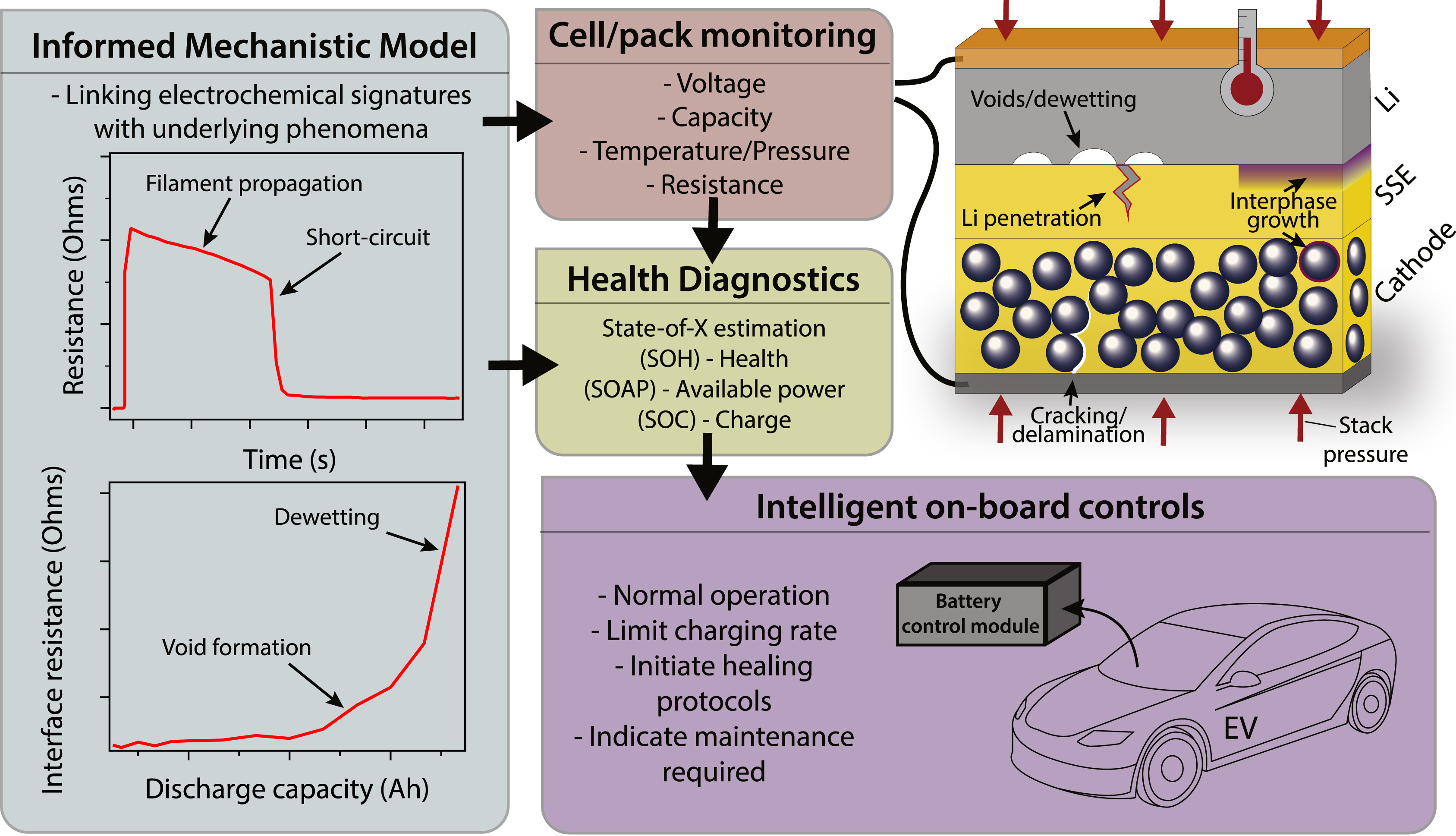}
    \caption{Diagram of a closed-loop SSB control scheme that couples mechanistic understanding with cell monitoring to optimize operation. Reproduced with permission from Ref. \cite{wangTransitioningSolidstateBatteries2021}.}
    \label{fig:wangTransitioning_integration}
\end{figure}

Wang \etal~\cite{wangTransitioningSolidstateBatteries2021} provide further details on the mechanical, thermal, and interfacial properties of SSBs and describe the role each of these properties play in transitioning an SSB concept from lab to market. Additionally, they emphasize the importance of a closed-loop design that also accounts for integration of battery cells with the technology they power. Besides allowing for control schemes which optimize performance and health of a SSB, cell monitoring can help communicate to the user necessary maintenance, and, in some cases, even perform diagnostics and reverse damage through subsequent charge/discharge cycles.~\cite{wangTransitioningSolidstateBatteries2021} For example, dendrite penetration and void growth have distinct signatures in the impedance profile of the cell, which can be used to better inform control strategies, as represented in Figure \ref{fig:wangTransitioning_integration}.

Li-free or anode-free SSBs entail a significant volume change during their operation, which presents challenges to assembling the cells together in modules and eventually into battery packs. Several SSB manufacturers have devised solutions to tackle this problem~\cite{shirasawa2012battery, hermann2014battery}. Such solutions are generally based on the use of a material or structure between cells to absorb the pressure generated from their volume expansion. The mechanical properties of inter-cell structure affects the chemomechanical response of the cells, thus influencing their performance. Battery packs made of SSBs also require control strategies that account for impacts that stack pressure and response of the inter-cell support structures can have on the performance of the battery pack~\cite{hermann2014battery}. Another prospect for using changes in the mechanical response of SSBs with charge and discharge is the use of change in strain~\cite{ningVisualizingPlatinginducedCracking2021}, as measured across the cell, to serve as a proxy for the state-of-charge of the cell. Further, across-cell strain measurements can be used to estimate changes in the charge-discharge characteristics over the lifetime of the cell, thereby providing an avenue to assess the change in Li inventory and hence the capacity of the cell, as well as estimate the rate of degradation of performance metrics.

\section{Conclusions \& Outlook}
To commercialize Li metal SSBs, a combination of criteria such as fast charge{/}discharge, high energy density, long cycle life and low cost need to be satisfied simultaneously. SSBs present unique challenges such as interfacial contact loss, Li metal penetration and cracking in solid electrolytes, void formation, etc. which occur under different conditions. Great progress has been made in identifying the microscale factors that contribute to failure. Stability criteria have been developed to identify different regimes involving the chemomechanical, transport and interfacial properties. Furthermore, similar to liquid electrolytes, various Li penetration morphologies into solid electrolytes have been observed for the same material even under the same operating conditions. A careful analysis of morphologies based on factors like material, microstructure and operating conditions needs to be developed. Moreover, the mechanisms of stabilization of SSBs based on glassy electrolytes like LiPON or coatings, which have been gaining attention, need to be investigated further. There may be potential in synthesis and processing techniques to emulate the success of glassy electrolytes. Finally, we note that, in general, obtaining sufficiently favorable overpotential changes solely due to mechanical stresses is very hard.  For example, a stress of 10 MPa, which is on the higher end of values for the yield stress of Li, contributes an overpotential of $\sim V_{\text{Li}} \sigma /F=1.3$ meV. Therefore, other phenomena to manipulate local current density like piezoelectric response~\cite{Liu2019preventing} and orientational ordering~\cite{Ahmad2020design} at the microscale may provide  additional driving forces to dynamically stabilize Li plating and stripping.

For applications that demand fast discharge, it is crucial to understand and address the kinetic driving forces behind pit formation. For such applications, alloying the Li anode has been attempted~\cite{krauskopfDiffusionLimitationLithium2019} but, in general, the use of pure Li metal anodes is almost always a requirement such as in anode free cells. Therefore, the focus lies on the effects that solid electrolytes, coatings or interlayers can have on this behavior. Design principles for such materials to mitigate pitting or dendrite penetration behavior will serve well to guide SSB development. Another area that requires further attention is understanding of the electrochemomechanical coupling at the space charge layer which dominates the interfacial interactions and possibly the resistance between the electrode and the electrolyte~\cite{becker-steinbergerStaticsDynamicsSpaceChargeLayers2021}. Overall, comprehending the nanoscale structure of these potentially multiphase interface regions and how they affect Li transport from a chemomechanical and charge transfer perspective is a necessary stepping stone for the mitigation of failure in Li metal SSBs.

\section*{Acknowledgements}
We thank Dr. Stephen Harris for helpful discussions. Authors acknowledge support from the Advanced Research Projects Agency-Energy (ARPA-E) under Grant DE-AR0000774.


 \bibliographystyle{elsarticle-ref}
 \bibliography{all_refs}

\providecommand{\noopsort}[1]{}\providecommand{\singleletter}[1]{#1}%
\begin{thebibliography}{100}
\expandafter\ifx\csname url\endcsname\relax
  \def\url#1{\texttt{#1}}\fi
\expandafter\ifx\csname urlprefix\endcsname\relax\def\urlprefix{URL }\fi
\expandafter\ifx\csname href\endcsname\relax
  \def\href#1#2{#2} \def\path#1{#1}\fi

\bibitem{manthiram2017solid}
A.~Manthiram, X.~Yu, S.~Wang,
  \href{http://dx.doi.org/10.1038/natrevmats.2016.103}{Lithium battery
  chemistries enabled by solid-state electrolytes}, Nat. Rev. Mater. 2 (2017)
  16103.

\bibitem{Robinson2014}
A.~L. Robinson, J.~Janek,
  \href{https://doi.org/10.1557/mrs.2014.285}{Solid-state batteries enter {EV}
  fray}, MRS Bull. 39~(12) (2014) 1046--1047.

\bibitem{janekSolidFutureBattery2016a}
J.~Janek, W.~G. Zeier, A solid future for battery development, Nat. Energy
  1~(9) (2016) 1--4.

\bibitem{fredericksPerformanceMetricsRequired2018}
W.~L. Fredericks, S.~Sripad, G.~C. Bower, V.~Viswanathan, Performance {{Metrics
  Required}} of {{Next}}-{{Generation Batteries}} to {{Electrify Vertical
  Takeoff}} and {{Landing}} ({{VTOL}}) {{Aircraft}}, ACS Energy Lett. 3~(12)
  (2018) 2989--2994.

\bibitem{moore2014aviation}
M.~D. Moore, \href{https://doi.org/10.2514/6.2014-0535}{Misconceptions of
  Electric Aircraft and their Emerging Aviation Markets}, AIAA SciTech Forum,
  American Institute of Aeronautics and Astronautics, 2014.

\bibitem{billsPerformanceMetricsRequired2020}
A.~Bills, S.~Sripad, W.~L. Fredericks, M.~Singh, V.~Viswanathan, Performance
  {{Metrics Required}} of {{Next}}-{{Generation Batteries}} to {{Electrify
  Commercial Aircraft}}, ACS Energy Lett. 5~(2) (2020) 663--668.

\bibitem{linDesignCationTransport2020a}
Y.-Y. Lin, A.~X.~B. Yong, W.~J. Gustafson, C.~N. Reedy, E.~Ertekin, J.~A.
  Krogstad, N.~H. Perry, Toward design of cation transport in solid-state
  battery electrolytes: {{Structure}}-dynamics relationships, Curr. Opin. Solid
  State Mater. Sci. 24~(6) (2020) 100875.

\bibitem{Xiao2019understanding}
Y.~Xiao, Y.~Wang, S.-H. Bo, J.~C. Kim, L.~J. Miara, G.~Ceder,
  \href{https://doi.org/10.1038/s41578-019-0157-5}{Understanding interface
  stability in solid-state batteries}, Nat. Rev. Mater. 5~(2) (2019) 105--126.

\bibitem{Wenzel2016direct}
S.~Wenzel, S.~Randau, T.~Leichtwei{\ss}, D.~A. Weber, J.~Sann, W.~G. Zeier,
  J.~Janek, \href{https://doi.org/10.1021/acs.chemmater.6b00610}{Direct
  observation of the interfacial instability of the fast ionic conductor
  li10gep2s12 at the lithium metal anode}, Chem. Mater. 28~(7) (2016)
  2400--2407.

\bibitem{Banerjee2020interfaces}
A.~Banerjee, X.~Wang, C.~Fang, E.~A. Wu, Y.~S. Meng,
  \href{https://doi.org/10.1021/acs.chemrev.0c00101}{Interfaces and interphases
  in all-solid-state batteries with inorganic solid electrolytes}, Chem. Rev.
  120~(14) (2020) 6878--6933.

\bibitem{Zhang2017electro}
W.~Zhang, D.~Schr\"{o}der, T.~Arlt, I.~Manke, R.~Koerver, R.~Pinedo, D.~A.
  Weber, J.~Sann, W.~G. Zeier, J.~Janek,
  \href{https://doi.org/10.1039/c7ta02730c}{(electro)chemical expansion during
  cycling: monitoring the pressure changes in operating solid-state lithium
  batteries}, J. Mater. Chem.A 5~(20) (2017) 9929--9936.

\bibitem{Yu2017accessing}
C.~Yu, S.~Ganapathy, E.~R.~H. van Eck, H.~Wang, S.~Basak, Z.~Li, M.~Wagemaker,
  \href{https://doi.org/10.1038/s41467-017-01187-y}{Accessing the bottleneck in
  all-solid state batteries, lithium-ion transport over the
  solid-electrolyte-electrode interface}, Nat. Commun. 8~(1) (2017) 1--9.

\bibitem{schnell2018all}
J.~Schnell, T.~G{\"u}nther, T.~Knoche, C.~Vieider, L.~K{\"o}hler, A.~Just,
  M.~Keller, S.~Passerini, G.~Reinhart, All-solid-state lithium-ion and lithium
  metal batteries--paving the way to large-scale production, J. Power Sources
  382 (2018) 160--175.

\bibitem{kamayaLithiumSuperionicConductor2011}
N.~Kamaya, K.~Homma, Y.~Yamakawa, M.~Hirayama, R.~Kanno, M.~Yonemura,
  T.~Kamiyama, Y.~Kato, S.~Hama, K.~Kawamoto, A.~Mitsui, A lithium superionic
  conductor, Nat. Mater. 10~(9) (2011) 682--686.

\bibitem{Seino2014}
Y.~Seino, T.~Ota, K.~Takada, A.~Hayashi, M.~Tatsumisago,
  \href{https://doi.org/10.1039/c3ee41655k}{A sulphide lithium super ion
  conductor is superior to liquid ion conductors for use in rechargeable
  batteries}, Energy Environ. Sci. 7~(2) (2014) 627--631.

\bibitem{Deiseroth2008}
H.-J. Deiseroth, S.-T. Kong, H.~Eckert, J.~Vannahme, C.~Reiner, T.~Zai{\ss},
  M.~Schlosser, \href{https://doi.org/10.1002/ange.200703900}{Li6ps5x: A class
  of crystalline li-rich solids with an unusually high {Li+} mobility}, Angew
  Chem-ger Edit 120~(4) (2008) 767--770.

\bibitem{Zhao2012superionic}
Y.~Zhao, L.~L. Daemen, \href{https://doi.org/10.1021/ja305709z}{Superionic
  conductivity in lithium-rich anti-perovskites}, J. Am. Chem. Soc. 134~(36)
  (2012) 15042--15047.

\bibitem{Zhang2013abinitio}
Y.~Zhang, Y.~Zhao, C.~Chen,
  \href{https://doi.org/10.1103/physrevb.87.134303}{Ab initiostudy of the
  stabilities of and mechanism of superionic transport in lithium-rich
  antiperovskites}, Phys. Rev. B 87~(13) (2013).

\bibitem{Thangadurai2003}
V.~Thangadurai, H.~Kaack, W.~J.~F. Weppner,
  \href{https://doi.org/10.1111/j.1151-2916.2003.tb03318.x}{Novel fast lithium
  ion conduction in garnet-type li5la3m2o12(m = nb, ta)}, J. Am. Ceram. Soc.
  86~(3) (2003) 437--440.

\bibitem{Murugan2007}
R.~Murugan, V.~Thangadurai, W.~Weppner,
  \href{https://doi.org/10.1002/anie.200701144}{Fast lithium ion conduction in
  garnet-type {Li7La3Zr2O12}}, Angew. Chem. - Int. Ed. 46~(41) (2007)
  7778--7781.

\bibitem{Kato2016-ssb}
Y.~Kato, S.~Hori, T.~Saito, K.~Suzuki, M.~Hirayama, A.~Mitsui, M.~Yonemura,
  H.~Iba, R.~Kanno, High-power all-solid-state batteries using sulfide
  superionic conductors, Nat. Energy 1 (2016) 16030.

\bibitem{Richards2015interface}
W.~D. Richards, L.~J. Miara, Y.~Wang, J.~C. Kim, G.~Ceder,
  \href{https://doi.org/10.1021/acs.chemmater.5b04082}{Interface stability in
  solid-state batteries}, Chem. Mater. 28~(1) (2015) 266--273.

\bibitem{Chen2019approaching}
R.~Chen, Q.~Li, X.~Yu, L.~Chen, H.~Li,
  \href{https://doi.org/10.1021/acs.chemrev.9b00268}{Approaching practically
  accessible solid-state batteries: Stability issues related to solid
  electrolytes and interfaces}, Chem. Rev. 120~(14) (2019) 6820--6877.

\bibitem{Ren2015direct}
Y.~Ren, Y.~Shen, Y.~Lin, C.-W. Nan, Direct observation of lithium dendrites
  inside garnet-type lithium-ion solid electrolyte, Electrochem. Commun. 57
  (2015) 27 -- 30.

\bibitem{porzMechanismLithiumMetal2017}
L.~Porz, T.~Swamy, B.~W. Sheldon, D.~Rettenwander, T.~Fr{\"o}mling, H.~L.
  Thaman, S.~Berendts, R.~Uecker, W.~C. Carter, Y.-M. Chiang, Mechanism of
  {{Lithium Metal Penetration}} through {{Inorganic Solid Electrolytes}}, Adv.
  Energy Mater. 7~(20) (2017) 1701003.

\bibitem{kasemchainanCriticalStrippingCurrent2019}
J.~Kasemchainan, S.~Zekoll, D.~Spencer~Jolly, Z.~Ning, G.~O. Hartley,
  J.~Marrow, P.~G. Bruce, Critical stripping current leads to dendrite
  formation on plating in lithium anode solid electrolyte cells, Nat. Mater.
  18~(10) (2019) 1105--1111.

\bibitem{huang2021manufacturing}
K.~J. Huang, G.~Ceder, E.~A. Olivetti,
  \href{https://doi.org/10.1016/j.joule.2020.12.001}{Manufacturing scalability
  implications of materials choice in inorganic solid-state batteries}, Joule
  5~(3) (2021) 564--580.

\bibitem{Lewis2019chemo}
J.~A. Lewis, J.~Tippens, F.~J.~Q. Cortes, M.~T. McDowell,
  \href{https://doi.org/10.1016/j.trechm.2019.06.013}{Chemo-mechanical
  challenges in solid-state batteries}, Trends Chem. 1~(9) (2019) 845--857.

\bibitem{zhaoReviewModelingElectrochemomechanics2019a}
Y.~Zhao, P.~Stein, Y.~Bai, M.~{Al-Siraj}, Y.~Yang, B.-X. Xu, A review on
  modeling of electro-chemo-mechanics in lithium-ion batteries, J. Power
  Sources 413 (2019) 259--283.

\bibitem{goyalNewFoundationsNewman2017}
P.~Goyal, C.~W. Monroe, New {{Foundations}} of {{Newman}}'s {{Theory}} for
  {{Solid Electrolytes}}: {{Thermodynamics}} and {{Transient Balances}}, J.
  Electrochem. Soc. 164~(11) (2017) E3647--E3660.

\bibitem{Monroe2004Effect}
C.~Monroe, J.~Newman, The effect of interfacial deformation on
  electrodeposition kinetics, J. Electrochem. Soc. 151~(6) (2004) A880--A886.

\bibitem{ganserExtendedFormulationButlerVolmer2019}
M.~Ganser, F.~E. Hildebrand, M.~Klinsmann, M.~Hanauer, M.~Kamlah, R.~M.
  McMeeking, An {{Extended Formulation}} of {{Butler}}-{{Volmer Electrochemical
  Reaction Kinetics Including}} the {{Influence}} of {{Mechanics}}, J.
  Electrochem. Soc. 166~(4) (2019) H167--H176.

\bibitem{Tikekar2016}
M.~D. Tikekar, L.~A. Archer, D.~L. Koch, Stabilizing electrodeposition in
  elastic solid electrolytes containing immobilized anions, Sci. Adv. 2~(7)
  (2016) 1600320.

\bibitem{Coleman1963}
B.~D. Coleman, W.~Noll, \href{https://doi.org/10.1007/bf01262690}{The
  thermodynamics of elastic materials with heat conduction and viscosity},
  Arch. Ration. Mech. Anal. 13~(1) (1963) 167--178.

\bibitem{bucciFormulationCoupledElectrochemical2016}
G.~Bucci, Y.-M. Chiang, W.~C. Carter, Formulation of the coupled
  electrochemical\textendash mechanical boundary-value problem, with
  applications to transport of multiple charged species, Acta Mater. 104 (2016)
  33--51.

\bibitem{ganserFiniteStrainElectrochemomechanical2019}
M.~Ganser, F.~E. Hildebrand, M.~Kamlah, R.~M. McMeeking, A finite strain
  electro-chemo-mechanical theory for ion transport with application to binary
  solid electrolytes, J. Mech. Phys. Solids 125 (2019) 681--713.

\bibitem{Monroe2005Impact}
C.~Monroe, J.~Newman, The impact of elastic deformation on deposition kinetics
  at lithium/polymer interfaces, J. Electrochem. Soc. 152~(2) (2005)
  A396--A404.

\bibitem{ahmad2017stability}
Z.~Ahmad, V.~Viswanathan, Stability of electrodeposition at solid-solid
  interfaces and implications for metal anodes, Phys. Rev. Lett. 119 (2017)
  056003.

\bibitem{baraiLithiumDendriteGrowth2017}
P.~Barai, K.~Higa, V.~Srinivasan, Lithium dendrite growth mechanisms in polymer
  electrolytes and prevention strategies, Phys. Chem. Chem. Phys. 19~(31)
  (2017) 20493--20505.

\bibitem{zhangPressureDrivenInterfaceEvolution2020}
X.~Zhang, Q.~J. Wang, K.~L. Harrison, S.~A. Roberts, S.~J. Harris,
  Pressure-{{Driven Interface Evolution}} in {{Solid}}-{{State Lithium Metal
  Batteries}}, Cell Rep. Phys. Sci. 1~(2) (2020) 100012.

\bibitem{Ahmad2020design}
Z.~Ahmad, Z.~Hong, V.~Viswanathan,
  \href{https://doi.org/10.1073/pnas.2008841117}{Design rules for liquid
  crystalline electrolytes for enabling dendrite-free lithium metal batteries},
  Proc. Natl. Acad. Sci. 117~(43) (2020) 26672--26680.

\bibitem{klinsmannDendriticCrackingSolid2019}
M.~Klinsmann, F.~E. Hildebrand, M.~Ganser, R.~M. McMeeking, Dendritic cracking
  in solid electrolytes driven by lithium insertion, J. Power Sources 442
  (2019) 227226.

\bibitem{pannikkat1999potential}
A.~Pannikkat, R.~Raj,
  \href{http://www.sciencedirect.com/science/article/pii/S1359645499002062}{Measurement
  of an electrical potential induced by normal stress applied to the interface
  of an ionic material at elevated temperatures}, Acta Mater. 47~(12) (1999)
  3423 -- 3431.

\bibitem{carmonaEffectMechanicalState2021}
E.~A. Carmona, M.~J. Wang, Y.~Song, J.~Sakamoto, P.~Albertus, The {{Effect}} of
  {{Mechanical State}} on the {{Equilibrium Potential}} of {{Alkali
  Metal}}/{{Ceramic Single}}-{{Ion Conductor Systems}}, Adv. Energy Mater.
  (2021) 2101355.

\bibitem{baraiMechanicalStressInduced2019}
P.~Barai, K.~Higa, A.~T. Ngo, L.~A. Curtiss, V.~Srinivasan, Mechanical {{Stress
  Induced Current Focusing}} and {{Fracture}} in {{Grain Boundaries}}, J.
  Electrochem. Soc. 166~(10) (2019) A1752.

\bibitem{McMeeking2019metal}
R.~M. McMeeking, M.~Ganser, M.~Klinsmann, F.~E. Hildebrand,
  \href{https://doi.org/10.1149/2.0221906jes}{Metal electrode surfaces can
  roughen despite the constraint of a stiff electrolyte}, J. Electrochem. Soc.
  166~(6) (2019) A984--A995.

\bibitem{lepageLithiumMechanicsRoles2019a}
W.~S. LePage, Y.~Chen, E.~Kazyak, K.-H. Chen, A.~J. Sanchez, A.~Poli, E.~M.
  Arruda, M.~D. Thouless, N.~P. Dasgupta, Lithium {{Mechanics}}: {{Roles}} of
  {{Strain Rate}} and {{Temperature}} and {{Implications}} for {{Lithium Metal
  Batteries}}, J. Electrochem. Soc. 166~(2) (2019) A89--A97.

\bibitem{herbertNanoindentationHighpurityVapor2018b}
E.~G. Herbert, S.~A. Hackney, N.~J. Dudney, P.~S. Phani, Nanoindentation of
  high-purity vapor deposited lithium films: {{The}} elastic modulus, Journal
  of Materials Research 33~(10) (2018) 1335--1346.

\bibitem{herbertNanoindentationHighpurityVapor2018}
E.~G. Herbert, S.~A. Hackney, V.~Thole, N.~J. Dudney, P.~S. Phani,
  Nanoindentation of high-purity vapor deposited lithium films: {{A}}
  mechanistic rationalization of diffusion-mediated flow, J. Mater. Res.
  33~(10) (2018) 1347--1360.

\bibitem{masiasElasticPlasticCreep2019}
A.~Masias, N.~Felten, R.~{Garcia-Mendez}, J.~Wolfenstine, J.~Sakamoto, Elastic,
  plastic, and creep mechanical properties of lithium metal, J. Mater. Sci.
  54~(3) (2019) 2585--2600.

\bibitem{xuLi2017}
C.~Xu, Z.~Ahmad, A.~Aryanfar, V.~Viswanathan, J.~R. Greer, Enhanced strength
  and temperature dependence of mechanical properties of li at small scales and
  its implications for li metal anodes, Proc. Natl. Acad. Sci. U.S.A. 114~(1)
  (2017) 57--61.

\bibitem{Slotwinski1969temperature}
T.~Slotwinski, J.~Trivisonno,
  \href{https://doi.org/10.1016/0022-3697(69)90386-2}{Temperature dependence of
  the elastic constants of single crystal lithium}, J. Phys. Chem. Solids
  30~(5) (1969) 1276--1278.

\bibitem{tariq2003li}
S.~Tariq, K.~Ammigan, P.~Hurh, R.~Schultz, P.~Liu, J.~Shang, Li material
  testing-fermilab antiproton source lithium collection lens, in: Proceedings
  of the 2003 particle accelerator conference, Vol.~3, IEEE, 2003, pp.
  1452--1454.

\bibitem{schultz2002lithium}
R.~P. Schultz, Lithium: Measurement of young's modulus and yield strength,
  Fermi National Accelerator Lab., Batavia, IL (US) (2002).

\bibitem{Greer2008comparing}
J.~R. Greer, C.~R. Weinberger, W.~Cai,
  \href{https://doi.org/10.1016/j.msea.2007.08.093}{Comparing the strength of
  f.c.c. and b.c.c. sub-micrometer pillars: Compression experiments and
  dislocation dynamics simulations}, Materials Science and Engineering: A
  493~(1-2) (2008) 21--25.

\bibitem{MinHan2013critical}
S.~M. Han, G.~Feng, J.~Y. Jung, H.~J. Jung, J.~R. Groves, W.~D. Nix, Y.~Cui,
  \href{https://doi.org/10.1063/1.4776658}{Critical-temperature/peierls-stress
  dependent size effects in body centered cubic nanopillars}, Appl. Phys. Lett.
  102~(4) (2013) 041910.

\bibitem{narayanLargeDeformationElastic2018}
S.~Narayan, L.~Anand, A large deformation elastic\textendash viscoplastic model
  for lithium, Extreme Mech. Lett. 24 (2018) 21--29.

\bibitem{wangNanoindentationStudyViscoplastic2017}
Y.~Wang, Y.-T. Cheng, A nanoindentation study of the viscoplastic behavior of
  pure lithium, Scr. Mater. 130 (2017) 191--195.

\bibitem{Messer1975}
R.~Messer, F.~Noack, \href{https://doi.org/10.1007/bf00883553}{Nuclear magnetic
  relaxation by self-diffusion in solid lithium:t 1-frequency dependence},
  Appl. Phys. 6~(1) (1975) 79--88.

\bibitem{Hao2018mesoscale}
F.~Hao, A.~Verma, P.~P. Mukherjee,
  \href{https://doi.org/10.1021/acsami.8b08796}{Mesoscale complexations in
  lithium electrodeposition}, ACS Appl. Mater. Interfaces 10~(31) (2018)
  26320--26327.

\bibitem{Jckle2018self}
M.~J\"{a}ckle, K.~Helmbrecht, M.~Smits, D.~Stottmeister, A.~Gro{\ss},
  \href{https://doi.org/10.1039/c8ee01448e}{Self-diffusion barriers: possible
  descriptors for dendrite growth in batteries?}, Energy Environ. Sci. 11~(12)
  (2018) 3400--3407.

\bibitem{chenLiMetalDeposition2020}
Y.~Chen, Z.~Wang, X.~Li, X.~Yao, C.~Wang, Y.~Li, W.~Xue, D.~Yu, S.~Y. Kim,
  F.~Yang, A.~Kushima, G.~Zhang, H.~Huang, N.~Wu, Y.-W. Mai, J.~B. Goodenough,
  J.~Li, Li metal deposition and stripping in a solid-state battery via
  {{Coble}} creep, Nature 578~(7794) (2020) 251--255.

\bibitem{Deng01012016}
Z.~Deng, Z.~Wang, I.-H. Chu, J.~Luo, S.~P. Ong, Elastic properties of alkali
  superionic conductor electrolytes from first principles calculations, J.
  Electrochem. Soc. 163~(2) (2016) A67--A74.

\bibitem{Ahmad16uncertainty}
Z.~Ahmad, V.~Viswanathan, Quantification of uncertainty in first-principles
  predicted mechanical properties of solids: Application to solid ion
  conductors, Phys. Rev. B 94 (2016) 064105.

\bibitem{yuElasticPropertiesSolid2016}
S.~Yu, R.~D. Schmidt, R.~{Garcia-Mendez}, E.~Herbert, N.~J. Dudney, J.~B.
  Wolfenstine, J.~Sakamoto, D.~J. Siegel, Elastic {{Properties}} of the {{Solid
  Electrolyte Li}}
  {\textsubscript{7}}{{La}}{\textsubscript{3}}{{Zr}}{\textsubscript{2}}{{O}}{\textsubscript{12}}
  ({{LLZO}}), Chem. Mater. 28~(1) (2016) 197--206.

\bibitem{Pugh1954}
S.~Pugh, Xcii. relations between the elastic moduli and the plastic properties
  of polycrystalline pure metals, Philos. Mag. Ser. 7 45~(367) (1954) 823--843.

\bibitem{garcia-mendezCorrelatingMacroAtomic2020}
R.~{Garcia-Mendez}, J.~G. Smith, J.~C. Neuefeind, D.~J. Siegel, J.~Sakamoto,
  Correlating {{Macro}} and {{Atomic Structure}} with {{Elastic Properties}}
  and {{Ionic Transport}} of {{Glassy Li2S}}-{{P2S5}} ({{LPS}}) {{Solid
  Electrolyte}} for {{Solid}}-{{State Li Metal Batteries}}, Adv. Energy Mater.
  10~(19) (2020) 2000335.

\bibitem{MullinsSekerka63}
W.~W. Mullins, R.~F. Sekerka, Morphological stability of a particle growing by
  diffusion or heat flow, J. Appl. Phys. 34~(2) (1963) 323--329.

\bibitem{MullinsSekerka64}
W.~W. Mullins, R.~F. Sekerka, Stability of a planar interface during
  solidification of a dilute binary alloy, J. Appl. Phys. 35~(2) (1964)
  444--451.

\bibitem{Asaro1972}
R.~J. Asaro, W.~A. Tiller, Interface morphology development during stress
  corrosion cracking: Part i. via surface diffusion, Metall. Trans. 3~(7)
  (1972) 1789--1796.

\bibitem{THEOFILIS2004}
V.~THEOFILIS, P.~W. DUCK, J.~OWEN,
  \href{https://doi.org/10.1017/s002211200400850x}{Viscous linear stability
  analysis of rectangular duct and cavity flows}, J. Fluid Mech. 505 (2004)
  249--286.

\bibitem{Pritzker1992}
M.~D. Pritzker, T.~Z. Fahidy,
  \href{https://doi.org/10.1016/0013-4686(92)80018-h}{Morphological stability
  of a planar metal electrode during potentiostatic electrodeposition and
  electrodissolution}, Electrochim. Acta 37~(1) (1992) 103--112.

\bibitem{Elezgaray1998}
J.~Elezgaray, C.~L{\'{e}}ger, F.~Argoul,
  \href{https://doi.org/10.1149/1.1838592}{Linear stability analysis of
  unsteady galvanostatic electrodeposition in the two-dimensional
  diffusion-limited regime}, J. Electrochem. Soc. 145~(6) (1998) 2016--2024.

\bibitem{ahmad2017-anisotropy}
Z.~Ahmad, V.~Viswanathan,
  \href{https://link.aps.org/doi/10.1103/PhysRevMaterials.1.055403}{Role of
  anisotropy in determining stability of electrodeposition at solid-solid
  interfaces}, Phys. Rev. Materials 1 (2017) 055403.

\bibitem{khooLinearStabilityAnalysis2019}
E.~Khoo, H.~Zhao, M.~Z. Bazant, Linear {{Stability Analysis}} of {{Transient
  Electrodeposition}} in {{Charged Porous Media}}: {{Suppression}} of
  {{Dendritic Growth}} by {{Surface Conduction}}, J. Electrochem. Soc. 166~(10)
  (2019) A2280.

\bibitem{Fu2020universal}
C.~Fu, V.~Venturi, J.~Kim, Z.~Ahmad, A.~W. Ells, V.~Viswanathan, B.~A. Helms,
  \href{https://doi.org/10.1038/s41563-020-0655-2}{Universal chemomechanical
  design rules for solid-ion conductors to prevent dendrite formation in
  lithium metal batteries}, Nat. Mater. 19~(7) (2020) 758--766.

\bibitem{mistryMolarVolumeMismatch2020}
A.~Mistry, P.~P. Mukherjee, Molar {{Volume Mismatch}}: {{A Malefactor}} for
  {{Irregular Metallic Electrodeposition}} with {{Solid Electrolytes}}, J.
  Electrochem. Soc. 167~(8) (2020) 082510.

\bibitem{baraiEffectInitialState2017}
P.~Barai, K.~Higa, V.~Srinivasan, Effect of {{Initial State}} of {{Lithium}} on
  the {{Propensity}} for {{Dendrite Formation}}: {{A Theoretical Study}}, J.
  Electrochem. Soc. 164~(2) (2017) A180--A189.

\bibitem{tuElectrodepositionMechanicalStability2020}
Q.~Tu, L.~{Barroso-Luque}, T.~Shi, G.~Ceder, Electrodeposition and {{Mechanical
  Stability}} at {{Lithium}}-{{Solid Electrolyte Interface}} during {{Plating}}
  in {{Solid}}-{{State Batteries}}, Cell Rep. Phys. Sci. 1~(7) (2020) 100106.

\bibitem{herbertMechanismsStressRelaxation2019}
E.~G. Herbert, N.~J. Dudney, M.~Rochow, V.~Thole, S.~A. Hackney, On the
  mechanisms of stress relaxation and intensification at the
  lithium/solid-state electrolyte interface, J. Mater. Res. 34~(21) (2019)
  3593--3616.

\bibitem{krauskopfFundamentalUnderstandingLithium2019}
T.~Krauskopf, H.~Hartmann, W.~G. Zeier, J.~Janek, Toward a {{Fundamental
  Understanding}} of the {{Lithium Metal Anode}} in {{Solid}}-{{State
  Batteries}}\textemdash{{An Electrochemo}}-{{Mechanical Study}} on the
  {{Garnet}}-{{Type Solid Electrolyte Li6}}.{{25Al0}}.{{25La3Zr2O12}}, ACS
  Appl. Mater. Interfaces 11~(15) (2019) 14463--14477.

\bibitem{zhangRethinkingHowExternal2019}
X.~Zhang, Q.~J. Wang, K.~L. Harrison, K.~Jungjohann, B.~L. Boyce, S.~A.
  Roberts, P.~M. Attia, S.~J. Harris, Rethinking {{How External Pressure Can
  Suppress Dendrites}} in {{Lithium Metal Batteries}}, J. Electrochem. Soc.
  166~(15) (2019) A3639.

\bibitem{douxPressureEffectsSulfide2020}
J.-M. Doux, Y.~Yang, D.~H.~S. Tan, H.~Nguyen, E.~A. Wu, X.~Wang, A.~Banerjee,
  Y.~S. Meng, Pressure effects on sulfide electrolytes for all solid-state
  batteries, J. Mater. Chem. A 8~(10) (2020) 5049--5055.

\bibitem{douxStackPressureConsiderations2020}
J.-M. Doux, H.~Nguyen, D.~H.~S. Tan, A.~Banerjee, X.~Wang, E.~A. Wu, C.~Jo,
  H.~Yang, Y.~S. Meng, Stack {{Pressure Considerations}} for
  {{Room}}-{{Temperature All}}-{{Solid}}-{{State Lithium Metal Batteries}},
  Adv. Energy Mater. 10~(1) (2020) 1903253.

\bibitem{kazyakLiPenetrationCeramic2020}
E.~Kazyak, R.~{Garcia-Mendez}, W.~S. LePage, A.~Sharafi, A.~L. Davis, A.~J.
  Sanchez, K.-H. Chen, C.~Haslam, J.~Sakamoto, N.~P. Dasgupta, Li
  {{Penetration}} in {{Ceramic Solid Electrolytes}}: {{Operando Microscopy
  Analysis}} of {{Morphology}}, {{Propagation}}, and {{Reversibility}}, Matter
  2~(4) (2020) 1025--1048.

\bibitem{vermaMicrostructurePressureDrivenElectrodeposition2021a}
A.~Verma, H.~Kawakami, H.~Wada, A.~Hirowatari, N.~Ikeda, Y.~Mizuno, T.~Kotaka,
  K.~Aotani, Y.~Tabuchi, P.~P. Mukherjee, Microstructure and
  {{Pressure}}-{{Driven Electrodeposition Stability}} in {{Solid}}-{{State
  Batteries}}, Cell Rep. Phys. Sci. 2~(1) (2021) 100301.

\bibitem{wangCorrelatingInterfaceResistance2018}
M.~Wang, J.~Sakamoto, Correlating the interface resistance and surface adhesion
  of the {{Li}} metal-solid electrolyte interface, J. Power Sources 377 (2018)
  7--11.

\bibitem{sharafiSurfaceChemistryMechanism2017}
A.~Sharafi, E.~Kazyak, A.~L. Davis, S.~Yu, T.~Thompson, D.~J. Siegel, N.~P.
  Dasgupta, J.~Sakamoto, Surface {{Chemistry Mechanism}} of {{Ultra}}-{{Low
  Interfacial Resistance}} in the {{Solid}}-{{State Electrolyte Li7La3Zr2O12}},
  Chem. Mater. 29~(18) (2017) 7961--7968.

\bibitem{hanNegatingInterfacialImpedance2017}
X.~Han, Y.~Gong, K.~K. Fu, X.~He, G.~T. Hitz, J.~Dai, A.~Pearse, B.~Liu,
  H.~Wang, G.~Rubloff, Y.~Mo, V.~Thangadurai, E.~D. Wachsman, L.~Hu, Negating
  interfacial impedance in garnet-based solid-state {{Li}} metal batteries,
  Nature Mater. 16~(5) (2017) 572--579.

\bibitem{Chen2017lithium}
L.~Chen, J.~G. Connell, A.~Nie, Z.~Huang, K.~R. Zavadil, K.~C. Klavetter,
  Y.~Yuan, S.~Sharifi-Asl, R.~Shahbazian-Yassar, J.~A. Libera, A.~U. Mane,
  J.~W. Elam, \href{https://doi.org/10.1039/c7ta03116e}{Lithium metal protected
  by atomic layer deposition metal oxide for high performance anodes}, J.
  Mater. Chem. A 5~(24) (2017) 12297--12309.

\bibitem{Zheng2020recent}
Z.-J. Zheng, H.~Ye, Z.-P. Guo,
  \href{https://doi.org/10.1002/advs.202002212}{Recent progress in designing
  stable composite lithium anodes with improved wettability}, Adv. Sci. 7~(22)
  (2020) 2002212.

\bibitem{Yu2018nanoflake}
B.~Yu, T.~Tao, S.~Mateti, S.~Lu, Y.~Chen,
  \href{https://doi.org/10.1002/adfm.201803023}{Nanoflake arrays of
  lithiophilic metal oxides for the ultra-stable anodes of lithium-metal
  batteries}, Advanced Functional Materials 28~(36) (2018) 1803023.

\bibitem{Wang2019tuning}
S.-H. Wang, J.~Yue, W.~Dong, T.-T. Zuo, J.-Y. Li, X.~Liu, X.-D. Zhang, L.~Liu,
  J.-L. Shi, Y.-X. Yin, Y.-G. Guo,
  \href{https://doi.org/10.1038/s41467-019-12938-4}{Tuning wettability of
  molten lithium via a chemical strategy for lithium metal anodes}, Nature
  Comm. 10~(1) (Oct. 2019).

\bibitem{Tao2020surface}
L.~Tao, A.~Hu, Z.~Yang, Z.~Xu, C.~E. Wall, A.~R. Esker, Z.~Zheng, F.~Lin,
  \href{https://doi.org/10.1002/adfm.202000585}{A surface chemistry approach to
  tailoring the hydrophilicity and lithiophilicity of carbon films for hosting
  high-performance lithium metal anodes}, Adv. Funct. Mater. 30~(31) (2020)
  2000585.

\bibitem{Sharafi2016characterizing}
A.~Sharafi, H.~M. Meyer, J.~Nanda, J.~Wolfenstine, J.~Sakamoto,
  \href{https://doi.org/10.1016/j.jpowsour.2015.10.053}{Characterizing the
  li{\textendash}li7la3zr2o12 interface stability and kinetics as a function of
  temperature and current density}, J. Power Sources 302 (2016) 135--139.

\bibitem{Cheng2015inter}
L.~Cheng, C.~H. Wu, A.~Jarry, W.~Chen, Y.~Ye, J.~Zhu, R.~Kostecki, K.~Persson,
  J.~Guo, M.~Salmeron, G.~Chen, M.~Doeff,
  \href{https://doi.org/10.1021/acsami.5b02528}{Interrelationships among grain
  size, surface composition, air stability, and interfacial resistance of
  al-substituted li7la3zr2o12 solid electrolytes}, {ACS} Applied Materials {\&}
  Interfaces 7~(32) (2015) 17649--17655.

\bibitem{ishiguroStabilityNbDopedCubic2013}
K.~Ishiguro, Y.~Nakata, M.~Matsui, I.~Uechi, Y.~Takeda, O.~Yamamoto,
  N.~Imanishi, Stability of {{Nb}}-{{Doped Cubic Li7La3Zr2O12}} with {{Lithium
  Metal}}, J. Electrochem. Soc. 160~(10) (2013) A1690.

\bibitem{sharafiCharacterizingLiLi7La3Zr2O122016}
A.~Sharafi, H.~M. Meyer, J.~Nanda, J.~Wolfenstine, J.~Sakamoto, Characterizing
  the {{Li}}\textendash{{Li7La3Zr2O12}} interface stability and kinetics as a
  function of temperature and current density, J. Power Sources 302 (2016)
  135--139.

\bibitem{Han2019high}
F.~Han, A.~S. Westover, J.~Yue, X.~Fan, F.~Wang, M.~Chi, D.~N. Leonard, N.~J.
  Dudney, H.~Wang, C.~Wang,
  \href{https://doi.org/10.1038/s41560-018-0312-z}{High electronic conductivity
  as the origin of lithium dendrite formation within solid electrolytes}, Nat.
  Energy 4~(3) (2019) 187--196.

\bibitem{Cheng2015effect}
L.~Cheng, W.~Chen, M.~Kunz, K.~Persson, N.~Tamura, G.~Chen, M.~Doeff,
  \href{https://doi.org/10.1021/am508111r}{Effect of surface microstructure on
  electrochemical performance of garnet solid electrolytes}, ACS Appl. Mater.
  Interfaces 7~(3) (2015) 2073--2081.

\bibitem{sharafiControllingCorrelatingEffect2017}
A.~Sharafi, C.~G. Haslam, R.~D. Kerns, J.~Wolfenstine, J.~Sakamoto, Controlling
  and correlating the effect of grain size with the mechanical and
  electrochemical properties of {{Li}} {\textsubscript{7}} {{La}}
  {\textsubscript{3}} {{Zr}} {\textsubscript{2}} {{O}} {\textsubscript{12}}
  solid-state electrolyte, J. Mater. Chem. A 5~(40) (2017) 21491--21504.

\bibitem{liDendriteNucleationLithiumconductive2019}
G.~Li, C.~W.~Monroe, Dendrite nucleation in lithium-conductive ceramics, Phys.
  Chem. Chem. Phys. 21~(36) (2019) 20354--20359.

\bibitem{Braun2015}
S.~Braun, C.~Yada, A.~Latz,
  \href{https://doi.org/10.1021/acs.jpcc.5b02679}{Thermodynamically consistent
  model for space-charge-layer formation in a solid electrolyte}, The Journal
  of Physical Chemistry C 119~(39) (2015) 22281--22288.

\bibitem{krauskopfDiffusionLimitationLithium2019}
T.~Krauskopf, B.~Mogwitz, C.~Rosenbach, W.~G. Zeier, J.~Janek, Diffusion
  {{Limitation}} of {{Lithium Metal}} and {{Li}}\textendash{{Mg Alloy Anodes}}
  on {{LLZO Type Solid Electrolytes}} as a {{Function}} of {{Temperature}} and
  {{Pressure}}, Adv. Energy Mater. 9~(44) (2019) 1902568.

\bibitem{Kinzer2021operando}
B.~Kinzer, A.~L. Davis, T.~Krauskopf, H.~Hartmann, W.~S. LePage, E.~Kazyak,
  J.~Janek, N.~P. Dasgupta, J.~Sakamoto,
  \href{https://doi.org/10.1016/j.matt.2021.04.016}{Operando analysis of the
  molten li$\vert${LLZO} interface: Understanding how the physical properties
  of li affect the critical current density}, Matter 4~(6) (2021) 1947--1961.

\bibitem{swamyLithiumMetalPenetration2018}
T.~Swamy, R.~Park, B.~W. Sheldon, D.~Rettenwander, L.~Porz, S.~Berendts,
  R.~Uecker, W.~C. Carter, Y.-M. Chiang, Lithium {{Metal Penetration Induced}}
  by {{Electrodeposition}} through {{Solid Electrolytes}}: {{Example}} in
  {{Single}}-{{Crystal Li}} {\textsubscript{6}} {{La}} {\textsubscript{3}}
  {{ZrTaO}} {\textsubscript{12}} {{Garnet}}, J. Electrochem. Soc. 165~(16)
  (2018) A3648--A3655.

\bibitem{bucciModelingInternalMechanical2017a}
G.~Bucci, T.~Swamy, Y.-M. Chiang, W.~C. Carter, Modeling of internal mechanical
  failure of all-solid-state batteries during electrochemical cycling, and
  implications for battery design, J. Mater. Chem. A 5~(36) (2017)
  19422--19430.

\bibitem{rajCurrentLimitDiagrams2017}
R.~Raj, J.~Wolfenstine, Current limit diagrams for dendrite formation in
  solid-state electrolytes for {{Li}}-ion batteries, J. Power Sources 343
  (2017) 119--126.

\bibitem{barroso-luqueAnalysisSolidStateElectrodepositionInduced2020}
L.~{Barroso-Luque}, Q.~Tu, G.~Ceder, An {{Analysis}} of {{Solid}}-{{State
  Electrodeposition}}-{{Induced Metal Plastic Flow}} and {{Predictions}} of
  {{Stress States}} in {{Solid Ionic Conductor Defects}}, J. Electrochem. Soc.
  167~(2) (2020) 020534.

\bibitem{xuEnhancedStrengthTemperature2017}
C.~Xu, Z.~Ahmad, A.~Aryanfar, V.~Viswanathan, J.~R. Greer, Enhanced strength
  and temperature dependence of mechanical properties of {{Li}} at small scales
  and its implications for {{Li}} metal anodes, Proc. Natl. Acad. Sci. U.S.A.
  114~(1) (2017) 57--61.

\bibitem{Cui2017-texturing}
F.~Shi, A.~Pei, A.~Vailionis, J.~Xie, B.~Liu, J.~Zhao, Y.~Gong, Y.~Cui,
  \href{http://www.pnas.org/content/114/46/12138.abstract}{Strong texturing of
  lithium metal in batteries}, Proc. Natl. Acad. Sci. U.S.A. 114~(46) (2017)
  12138--12143.

\bibitem{ningVisualizingPlatinginducedCracking2021}
Z.~Ning, D.~S. Jolly, G.~Li, R.~De~Meyere, S.~D. Pu, Y.~Chen, J.~Kasemchainan,
  J.~Ihli, C.~Gong, B.~Liu, D.~L.~R. Melvin, A.~Bonnin, O.~Magdysyuk,
  P.~Adamson, G.~O. Hartley, C.~W. Monroe, T.~J. Marrow, P.~G. Bruce,
  Visualizing plating-induced cracking in lithium-anode solid-electrolyte
  cells, Nat. Mater. (2021) 1--9.

\bibitem{Qi2020anew}
Y.~Qi, C.~Ban, S.~J. Harris,
  \href{https://doi.org/10.1016/j.joule.2020.10.009}{A new general paradigm for
  understanding and preventing li metal penetration through solid
  electrolytes}, Joule 4~(12) (2020) 2599--2608.

\bibitem{varley2017polyborane}
J.~B. Varley, K.~Kweon, P.~Mehta, P.~Shea, T.~W. Heo, T.~J. Udovic, V.~Stavila,
  B.~C. Wood,
  \href{https://doi.org/10.1021/acsenergylett.6b00620}{Understanding ionic
  conductivity trends in polyborane solid electrolytes from ab initio molecular
  dynamics}, ACS Energy Lett. 2~(1) (2017) 250--255.

\bibitem{Ye2021dynamic}
L.~Ye, X.~Li, \href{https://doi.org/10.1038/s41586-021-03486-3}{A dynamic
  stability design strategy for lithium metal solid state batteries}, Nature
  593~(7858) (2021) 218--222.

\bibitem{epsteinConsiderationsReducingAviation2019}
A.~H. Epstein, S.~M. O'Flarity, Considerations for {{Reducing Aviation}}'s
  {{CO2}} with {{Aircraft Electric Propulsion}}, J. Propul. Power. 35~(3)
  (2019) 572--582.

\bibitem{wangCharacterizingLiSolidElectrolyteInterface2019}
M.~J. Wang, R.~Choudhury, J.~Sakamoto, Characterizing the
  {{Li}}-{{Solid}}-{{Electrolyte Interface Dynamics}} as a {{Function}} of
  {{Stack Pressure}} and {{Current Density}}, Joule 3~(9) (2019) 2165--2178.

\bibitem{woodDendritesPitsUntangling2016}
K.~N. Wood, E.~Kazyak, A.~F. Chadwick, K.-H. Chen, J.-G. Zhang, K.~Thornton,
  N.~P. Dasgupta, Dendrites and {{Pits}}: {{Untangling}} the {{Complex
  Behavior}} of {{Lithium Metal Anodes}} through {{Operando Video Microscopy}},
  ACS Cent. Sci. 2~(11) (2016) 790--801.

\bibitem{cohenMicromorphologicalStudiesLithium2000}
Y.~S. Cohen, Y.~Cohen, D.~Aurbach, Micromorphological {{Studies}} of {{Lithium
  Electrodes}} in {{Alkyl Carbonate Solutions Using}} in {{Situ Atomic Force
  Microscopy}}, J. Phys. Chem. B 104~(51) (2000) 12282--12291.

\bibitem{shiLithiumMetalStripping2018}
F.~Shi, A.~Pei, D.~T. Boyle, J.~Xie, X.~Yu, X.~Zhang, Y.~Cui, Lithium metal
  stripping beneath the solid electrolyte interphase, Proc. Natl. Acad. Sci.
  U.S.A. 115~(34) (2018) 8529--8534.

\bibitem{lewisLinkingVoidInterphase2021}
J.~A. Lewis, F.~J.~Q. Cortes, Y.~Liu, J.~C. Miers, A.~Verma, B.~S. Vishnugopi,
  J.~Tippens, D.~Prakash, T.~S. Marchese, S.~Y. Han, C.~Lee, P.~P. Shetty,
  H.-W. Lee, P.~Shevchenko, F.~De~Carlo, C.~Saldana, P.~P. Mukherjee, M.~T.
  McDowell, Linking void and interphase evolution to electrochemistry in
  solid-state batteries using operando {{X}}-ray tomography, Nat. Mater. 20~(4)
  (2021) 503--510.

\bibitem{sanchezLithiumStrippingAnisotropic2021}
A.~J. Sanchez, E.~Kazyak, Y.~Chen, J.~Lasso, N.~P. Dasgupta, Lithium stripping:
  Anisotropic evolution and faceting of pits revealed by operando 3-{{D}}
  microscopy, J. Mater. Chem. A (Jul. 2021).

\bibitem{sanchezPlanViewOperandoVideo2020}
A.~J. Sanchez, E.~Kazyak, Y.~Chen, K.-H. Chen, E.~R. Pattison, N.~P. Dasgupta,
  Plan-{{View Operando Video Microscopy}} of {{Li Metal Anodes}}:
  {{Identifying}} the {{Coupled Relationships}} among {{Nucleation}},
  {{Morphology}}, and {{Reversibility}}, ACS Energy Lett. 5~(3) (2020)
  994--1004.

\bibitem{venturiThermodynamicsLithiumStripping2021}
V.~Venturi, V.~Viswanathan, Thermodynamics of {{Lithium Stripping}} and
  {{Limits}} for {{Fast Discharge}} in {{Lithium Metal Batteries}},
  arXiv:2103.03921 (2021).

\bibitem{yangMaintainingFlatLi2021}
C.-T. Yang, Y.~Qi, Maintaining a {{Flat Li Surface}} during the {{Li Stripping
  Process}} via {{Interface Design}}, Chem. Mater. 33~(8) (2021) 2814--2823.

\bibitem{sudoInterfaceBehaviorGarnettype2014}
R.~Sudo, Y.~Nakata, K.~Ishiguro, M.~Matsui, A.~Hirano, Y.~Takeda, O.~Yamamoto,
  N.~Imanishi, Interface behavior between garnet-type lithium-conducting solid
  electrolyte and lithium metal, Solid State Ionics 262 (2014) 151--154.

\bibitem{renDirectObservationLithium2015}
Y.~Ren, Y.~Shen, Y.~Lin, C.-W. Nan, Direct observation of lithium dendrites
  inside garnet-type lithium-ion solid electrolyte, Electrochem. Commun. 57
  (2015) 27--30.

\bibitem{chengIntergranularLiMetal2017}
E.~J. Cheng, A.~Sharafi, J.~Sakamoto, Intergranular {{Li}} metal propagation
  through polycrystalline {{Li6}}.{{25Al0}}.{{25La3Zr2O12}} ceramic
  electrolyte, Electrochim. Acta 223 (2017) 85--91.

\bibitem{tsaiLi7La3Zr2O12InterfaceModification2016}
C.-L. Tsai, V.~Roddatis, C.~V. Chandran, Q.~Ma, S.~Uhlenbruck, M.~Bram,
  P.~Heitjans, O.~Guillon, {{Li7La3Zr2O12 Interface Modification}} for {{Li
  Dendrite Prevention}}, ACS Appl. Mater. Interfaces 8~(16) (2016)
  10617--10626.

\bibitem{yuGrainBoundarySoftening2018}
S.~Yu, D.~J. Siegel, Grain {{Boundary Softening}}: {{A Potential Mechanism}}
  for {{Lithium Metal Penetration}} through {{Stiff Solid Electrolytes}}, ACS
  Appl. Mater. Interfaces 10~(44) (2018) 38151--38158.

\bibitem{baraiRoleLocalInhomogeneities2020}
P.~Barai, A.~T. Ngo, B.~Narayanan, K.~Higa, L.~A. Curtiss, V.~Srinivasan, The
  {{Role}} of {{Local Inhomogeneities}} on {{Dendrite Growth}} in
  {{LLZO}}-{{Based Solid Electrolytes}}, J. Electrochem. Soc. 167~(10) (2020)
  100537.

\bibitem{yuGrainBoundaryContributions2017}
S.~Yu, D.~J. Siegel, Grain {{Boundary Contributions}} to {{Li}}-{{Ion
  Transport}} in the {{Solid Electrolyte Li7La3Zr2O12}} ({{LLZO}}), Chem.
  Mater. 29~(22) (2017) 9639--9647.

\bibitem{dixitTortuosityEffectsGarnetType2019}
M.~B. Dixit, M.~Regala, F.~Shen, X.~Xiao, K.~B. Hatzell, Tortuosity {{Effects}}
  in {{Garnet}}-{{Type Li}} {\textsubscript{7}} {{La}} {\textsubscript{3}}
  {{Zr}} {\textsubscript{2}} {{O}} {\textsubscript{12}} {{Solid Electrolytes}},
  ACS Appl. Mater. Interfaces 11~(2) (2019) 2022--2030.

\bibitem{shenEffectPoreConnectivity2018}
F.~Shen, M.~B. Dixit, X.~Xiao, K.~B. Hatzell, Effect of {{Pore Connectivity}}
  on {{Li Dendrite Propagation}} within {{LLZO Electrolytes Observed}} with
  {{Synchrotron X}}-ray {{Tomography}}, ACS Energy Lett. 3~(4) (2018)
  1056--1061.

\bibitem{liSolidElectrolyteKey2015}
J.~Li, C.~Ma, M.~Chi, C.~Liang, N.~J. Dudney, Solid {{Electrolyte}}: The
  {{Key}} for {{High}}-{{Voltage Lithium Batteries}}, Adv. Energy Mater. 5~(4)
  (2015) 1401408.

\bibitem{Bates1993}
J.~Bates, N.~Dudney, G.~Gruzalski, R.~Zuhr, A.~Choudhury, C.~Luck,
  J.~Robertson, \href{https://doi.org/10.1016/0378-7753(93)80106-y}{Fabrication
  and characterization of amorphous lithium electrolyte thin films and
  rechargeable thin-film batteries}, J. Power Sources 43~(1-3) (1993) 103--110.

\bibitem{Bates2000}
J.~Bates, \href{https://doi.org/10.1016/s0167-2738(00)00327-1}{Thin-film
  lithium and lithium-ion batteries}, Solid State Ionics 135~(1-4) (2000)
  33--45.

\bibitem{albertusStatusChallengesEnabling2018}
P.~Albertus, S.~Babinec, S.~Litzelman, A.~Newman, Status and challenges in
  enabling the lithium metal electrode for high-energy and low-cost
  rechargeable batteries, Nat. Energy 3~(1) (2018) 16--21.

\bibitem{westoverDepositionConfinementLi2019}
A.~S. Westover, N.~J. Dudney, R.~L. Sacci, S.~Kalnaus, Deposition and
  {{Confinement}} of {{Li Metal}} along an {{Artificial
  Lipon}}\textendash{{Lipon Interface}}, ACS Energy Lett. 4~(3) (2019)
  651--655.

\bibitem{sastreBlockingLithiumDendrite2021}
J.~Sastre, M.~H. Futscher, L.~Pompizi, A.~Aribia, A.~Priebe, J.~Overbeck,
  M.~Stiefel, A.~N. Tiwari, Y.~E. Romanyuk, Blocking lithium dendrite growth in
  solid-state batteries with an ultrathin amorphous {{Li}}-{{La}}-{{Zr}}-{{O}}
  solid electrolyte, Commun. Mater. 2~(1) (2021) 76.

\bibitem{zhengGrainBoundaryModification2021}
C.~Zheng, Y.~Ruan, J.~Su, Z.~Song, T.~Xiu, J.~Jin, M.~E. Badding, Z.~Wen, Grain
  boundary modification in garnet electrolyte to suppress lithium dendrite
  growth, Chem. Eng. J. 411 (2021) 128508.

\bibitem{huangEnhancedPerformanceLi2020}
Z.~Huang, L.~Chen, B.~Huang, B.~Xu, G.~Shao, H.~Wang, Y.~Li, C.-A. Wang,
  Enhanced {{Performance}} of {{Li}} {\textsubscript{6.4}} {{La}}
  {\textsubscript{3}} {{Zr}} {\textsubscript{1.4}} {{Ta}} {\textsubscript{0.6}}
  {{O}} {\textsubscript{12}} {{Solid Electrolyte}} by the {{Regulation}} of
  {{Grain}} and {{Grain Boundary Phases}}, ACS Appl. Mater. Interfaces 12~(50)
  (2020) 56118--56125.

\bibitem{hanHighElectronicConductivity2019}
F.~Han, A.~S. Westover, J.~Yue, X.~Fan, F.~Wang, M.~Chi, D.~N. Leonard, N.~J.
  Dudney, H.~Wang, C.~Wang, High electronic conductivity as the origin of
  lithium dendrite formation within solid electrolytes, Nat. Energy 4~(3) (Jan.
  2019).

\bibitem{aguesseInvestigatingDendriticGrowth2017}
F.~Aguesse, W.~Manalastas, L.~Buannic, J.~M. {Lopez del Amo}, G.~Singh,
  A.~Llord{\'e}s, J.~Kilner, Investigating the {{Dendritic Growth}} during
  {{Full Cell Cycling}} of {{Garnet Electrolyte}} in {{Direct Contact}} with
  {{Li Metal}}, ACS Appl. Mater. Interfaces 9~(4) (2017) 3808--3816.

\bibitem{songProbingOriginElectronic2019}
Y.~Song, L.~Yang, L.~Tao, Q.~Zhao, Z.~Wang, Y.~Cui, H.~Liu, Y.~Lin, F.~Pan,
  Probing into the origin of an electronic conductivity surge in a garnet
  solid-state electrolyte, J. Mater. Chem. A 7~(40) (2019) 22898--22902.

\bibitem{liuLocalElectronicStructure2021}
X.~Liu, R.~{Garcia-Mendez}, A.~R. Lupini, Y.~Cheng, Z.~D. Hood, F.~Han,
  A.~Sharafi, J.~C. Idrobo, N.~J. Dudney, C.~Wang, C.~Ma, J.~Sakamoto, M.~Chi,
  Local electronic structure variation resulting in {{Li}} `filament' formation
  within solid electrolytes, Nat. Mater. (2021) 1--6.

\bibitem{tianInterfacialElectronicProperties2019}
H.-K. Tian, Z.~Liu, Y.~Ji, L.-Q. Chen, Y.~Qi, Interfacial {{Electronic
  Properties Dictate Li Dendrite Growth}} in {{Solid Electrolytes}}, Chem.
  Mater. 31~(18) (2019) 7351--7359.

\bibitem{fengImpactElectronicProperties2021}
M.~Feng, J.~Pan, Y.~Qi, Impact of {{Electronic Properties}} of {{Grain
  Boundaries}} on the {{Solid Electrolyte Interphases}} ({{SEIs}}) in
  {{Li}}-ion {{Batteries}}, J. Phys. Chem. C (2021).

\bibitem{Ahmad2021interfaces}
Z.~Ahmad, V.~Venturi, H.~Hafiz, V.~Viswanathan,
  \href{https://doi.org/10.1021/acs.jpcc.1c00867}{Interfaces in solid
  electrolyte interphase: Implications for lithium-ion batteries}, J. Phys.
  Chem. C 125~(21) (2021) 11301--11309.

\bibitem{duffner2021post}
F.~Duffner, N.~Kronemeyer, J.~T{\"u}bke, J.~Leker, M.~Winter, R.~Schmuch,
  Post-lithium-ion battery cell production and its compatibility with
  lithium-ion cell production infrastructure, Nat. Energy 6~(2) (2021)
  123--134.

\bibitem{hatzell2021prospects}
K.~B. Hatzell, Y.~Zheng, Prospects on large-scale manufacturing of solid state
  batteries, MRS Energy Sustain. (2021) 1--7.

\bibitem{wang2020enabling}
M.~J. Wang, E.~Carmona, A.~Gupta, P.~Albertus, J.~Sakamoto, Enabling
  “lithium-free” manufacturing of pure lithium metal solid-state batteries
  through in situ plating, Nat. Commun. 11~(1) (2020) 1--9.

\bibitem{wangTransitioningSolidstateBatteries2021}
M.~J. Wang, E.~Kazyak, N.~P. Dasgupta, J.~Sakamoto, Transitioning solid-state
  batteries from lab to market: {{Linking}} electro-chemo-mechanics with
  practical considerations, Joule 5~(6) (2021) 1371--1390.

\bibitem{schnell2019prospects}
J.~Schnell, F.~Tietz, C.~Singer, A.~Hofer, N.~Billot, G.~Reinhart, Prospects of
  production technologies and manufacturing costs of oxide-based
  all-solid-state lithium batteries, Energy Environ. Sci. 12~(6) (2019)
  1818--1833.

\bibitem{placke2017lithium}
T.~Placke, R.~Kloepsch, S.~D\"{u}hnen, M.~Winter,
  \href{https://doi.org/10.1007/s10008-017-3610-7}{Lithium ion, lithium metal,
  and alternative rechargeable battery technologies: the odyssey for high
  energy density}, J. Solid State Electrochem. 21~(7) (2017) 1939--1964.

\bibitem{wang2020electrochemical}
M.~Wang, J.~Sakamoto,
  \href{https://doi.org/10.1149/ma2020-012293mtgabs}{Electrochemical formation
  of li metal anodes to enable li-free manufacturing of solid-state batteries},
  {ECS} Meeting Abstracts {MA}2020-01~(2) (2020) 293--293.

\bibitem{balaish2021processing}
M.~Balaish, J.~C. Gonzalez-Rosillo, K.~J. Kim, Y.~Zhu, Z.~D. Hood, J.~L. Rupp,
  Processing thin but robust electrolytes for solid-state batteries, Nat.
  Energy 6~(3) (2021) 227--239.

\bibitem{shirasawa2012battery}
A.~Shirasawa, H.~Teranishi, H.~Imai, Battery system and battery structure, {US}
  Patent App. 13/372,980 (Aug.~16 2012).

\bibitem{hermann2014battery}
W.~A. Hermann, Battery control systems, {US} Patent App. 14/038,763 (Apr.~3
  2014).

\bibitem{Liu2019preventing}
G.~Liu, D.~Wang, J.~Zhang, A.~Kim, W.~Lu,
  \href{https://doi.org/10.1021/acsmaterialslett.9b00289}{Preventing dendrite
  growth by a soft piezoelectric material}, ACS Mat. Lett. 1~(5) (2019)
  498--505.

\bibitem{becker-steinbergerStaticsDynamicsSpaceChargeLayers2021}
K.~{Becker-Steinberger}, S.~Schardt, B.~Horstmann, A.~Latz, Statics and
  {{Dynamics}} of {{Space}}-{{Charge}}-{{Layers}} in {{Polarized Inorganic
  Solid Electrolytes}}, arXiv:2101.10294 (2021).

\end{thebibliography}





\end{document}